\documentclass[final,3p,times]{elsarticle}

\usepackage{amsmath}
\usepackage{amssymb}
\usepackage[section]{placeins}
\usepackage{subcaption}
\usepackage{textcomp, gensymb}
\usepackage{verbatim}
\usepackage{array}         
\usepackage{tabularx}      
\usepackage{booktabs}      
\usepackage{soul}          
\usepackage{siunitx}        

\DeclareMathAlphabet{\mathcal}{OMS}{cmsy}{m}{n}

\journal{Journal of Computational Physics}

\begin{document}

\begin{frontmatter}



\title{Enhancing semi-resolved CFD-DEM for dilute to dense particle-fluid systems: A point cloud based, two-step mapping strategy via coarse graining}

\author[label1]{Yuxiang Liu}
\author[label1,label2]{Lu Jing\corref{cor1}}
\author[label2]{Xudong Fu}
\author[label3]{Huabin Shi}

\affiliation[label1]{organization={Institute for Ocean Engineering, Shenzhen International Graduate School, Tsinghua University},
            city={Shenzhen},
            country={China}}
\affiliation[label2]{organization={State Key Laboratory of Hydroscience and Engineering, Department of Hydraulic Engineering, Tsinghua University},
            city={Beijing},
            country={China}}
\affiliation[label3]{organization={State Key Laboratory of Internet of Things for Smart City and Department of Ocean Science and Technology, University of Macau},
            city={Macau},
            country={China}
            }
            
\cortext[cor1]{Corresponding author: lujing@sz.tsinghua.edu.cn}

\begin{abstract}
Computational fluid dynamics and discrete element method (CFD-DEM) coupling is an efficient and powerful tool to simulate particle-fluid systems. However, current volume-averaged CFD-DEM relying on direct grid-based mapping between the fluid and particle phases can exhibit a strong dependence on the fluid grid resolution, becoming unstable as particles move across fluid grids, and can fail to capture pore fluid pressure effects in very dense granular systems. Here we propose a two-step mapping CFD-DEM which uses a point-based coarse graining technique for intermediate smoothing to overcome these limitations. The discrete particles are first converted into smooth, coarse-grained continuum fields via a multi-layer Fibonacci point cloud, independent of the fluid grids. Then, accurate coupling is achieved between the coarse-grained, point cloud fields and the fluid grid-based variables. The algorithm is validated in various configurations, including weight allocation of a static particle on one-dimensional grids and a falling particle on two-dimensional grids, sedimentation of a sphere in a viscous fluid, size-bidisperse fluidized beds, Ergun's pressure drop test, and immersed granular column collapse. The proposed CFD-DEM represents a novel strategy to accurately simulate fluid-particle interactions for a wide range of grid-to-particle size ratios and solid concentrations, which is of potential use in many industrial and geophysical applications.
\end{abstract}




\end{frontmatter}


\section{Introduction} \label{sec:intro}

Particle-fluid systems are ubiquitous in natural phenomena and industrial processes, including sediment transport, debris flows, submarine landslides, slurry transport, and deep sea mining~\cite{talling2007onset,deal2023grain,peacock2023fluid}. Beyond the classic challenges in granular media, coupling of the solid particles with a fluid phase significantly complicates the problem due to granular dilatation and contraction~\cite{shi2021theoretical}, contact lubrication~\cite{topin2012collapse,ishikawa2022lubrication,dance2003incorporation}, rheology~\cite{boyer2011unifying,blais2016development}, and pore-pressure feedback~\cite{rondon2011granular,pailha2008initiation}. Understanding such phenomena requires accurate yet efficient numerical simulations of fluid-particle interactions for a wide range of particle sizes and solid concentrations. However, a robust method that can handle these scenarios, particularly for very dense granular systems where pore-scale effects dominate, is still missing.

Numerical methods for particle-fluid coupling can be broadly classified into Eulerian-Eulerian and Eulerian-Lagrangian categories. The former considers solid particles as a continuum phase using empirical granular rheology~\cite{jop2006constitutive,guazzelli2018rheology} or non-Newtonian viscosity models~\cite{sun2023two,malkus1990dynamics}, whereas the latter exploits a discrete description at the particle scale to avoid the use of empirical closures for the solid phase. A typical Eulerian-Lagrangian framework is the coupled computational fluid dynamics and discrete element method (CFD-DEM), which can be further divided into resolved and unresolved approaches depending on how well the fluid dynamics is resolved with respect to the particle scale. Resolved CFD-DEM approaches determine the particle-fluid interaction forces via integral along the interface. The interaction force can be treated through the body-fitted mesh method~\cite{hu2001direct}, immersed boundary method~\cite{peskin1977numerical,feng2005proteus}, or fictitious domain method~\cite{yu2006fictitious,yu2007direct}. The fluid mesh needs to be very fine (typically at least ten times smaller than the particle size) to achieve accurate particle-fluid interactions~\cite{uhlmann2005immersed}, which is extremely expensive. On the contrary, unresolved CFD-DEM, first proposed by Tsuji~\cite{tsuji1992lagrangian,tsuji1993discrete}, is more computationally affordable and hence widely used. In this framework, the fluid phase is solved by the volume-averaged Navier-Stokes (N-S) equations in CFD, particle motions are solved by Newton's second law in DEM, and particle-fluid interactions (e.g., drag force, virtual mass force, and lift force) are modeled through semi-empirical formulas. Time-consuming computations along solid-fluid interfaces are avoided, thus significantly improving the computational efficiency. 

\begin{figure}[htbp]
        \centerline{\includegraphics[height=8cm, keepaspectratio=true]{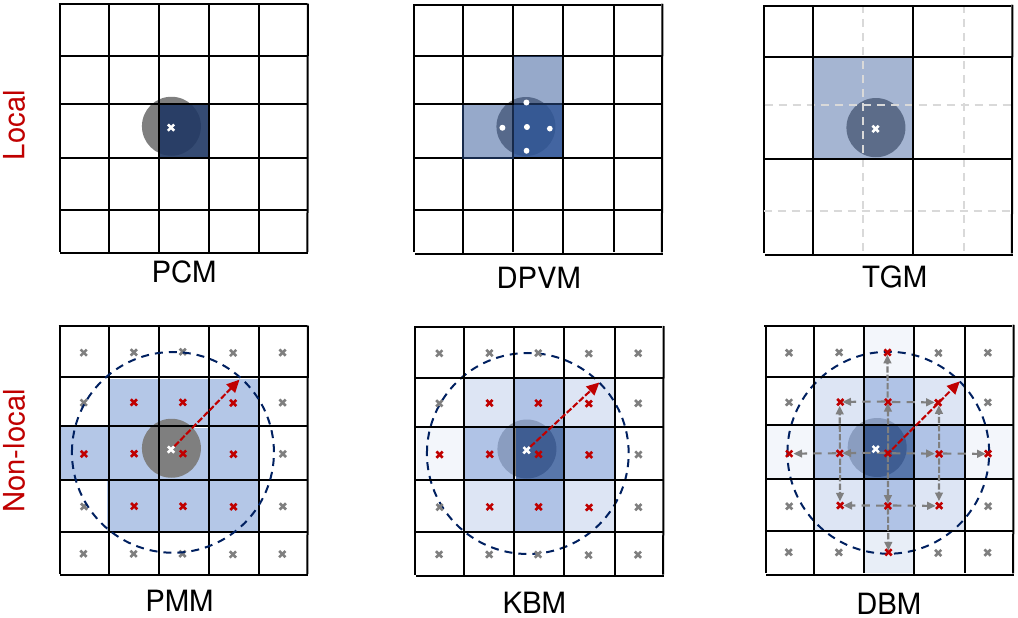}}
    \caption{\label{fig_compareAllmethod} Local and non-local mapping strategies for CFD-DEM. Crosses represent the centers of particles or fluid grids, while dots are point clouds used to identify portions of a particle. Arrows in the non-local methods indicate the searching areas. Darker shading indicates a higher weight when the particle information is mapped to the grid.}
\end{figure}

This paper focuses on unresolved CFD-DEM and its semi-resolved variants, whose computational accuracy relies heavily on how information (e.g., volume fraction, velocity, and interaction forces) is mapped between CFD grids and DEM particles. Various mapping strategies for CFD-DEM are illustrated in Fig.~\ref{fig_compareAllmethod}. The particle centroid model (PCM) is the most straightforward one~\cite{crowe1977particle}, which simply assigns a particle to a fluid grid based on where the particle centroid is located. This primitive method is known to cause erroneous results for fine grids~\cite{jing2016extended}. The divided particle volume method (DPVM) and satellite point methods are developed to better assign portions of a particle to their respective containing grids~\cite{wu2009three,peng2014influence,latzel2000macroscopic,gao2021development,wang2022super}, but they still rely on the ``local''
grids geometrically occupied by the particle, leading to unphysically high gradients of field quantities and, hence, numerical oscillations in fine grids~\cite{wang2019semi,zhang2023calculation,su2025novel}. 
Apart from particle partitioning methods, mapping techniques based on grid partitioning are also developed, such as the two grid method (TGM)~\cite{deb2013novel} and Voronoi grid-based method~\cite{che2021novel}, which use an alternative grid to enhance the calculation of volume fractions and particle-fluid interactions.

Unlike local coupling methods that only consider the grids where particles or their portions are located, so-called semi-resolved CFD-DEM approaches with a certain non-local smoothing techniques have been proposed to avoid high gradients in the computed flow fields (see Fig.~\ref{fig_compareAllmethod}, lower pannel). Link et al.~\cite{link2005flow} and Jing et al.~\cite{jing2016extended} employed the porous mapping method (PMM) to consider particle-fluid interactions in a larger region than the local CFD grids (i.e., a porous cube or sphere), with an equal weight, which can be regarded as the precursor of kernel-based methods (KBM). 
In KBM, a kernel function is used to weight the contribution of particles to a physical field based on particle-to-grid-center distances~\cite{kitagawa2001two}. 
Wang et al.~\cite{wang2019semi,wang2020semi,wang2021determination} developed a semi-resolved CFD-DEM approach to correct the drag force calculation using KBM. This method provides more accurate and stable results when the particle size is much larger than the fluid grids. Zhu et al.~\cite{zhu2022semi} extended this method for the transport of scalar fields (e.g., temperature).
Considering the stability of both fluid and solid phases, Chen and Zhang~\cite{chen2022semi} proposed a two-way domain expansion method with two cut-off distances based on particles and fluid grids. Zhang et al.~\cite{zhang2021optimized} proposed a virtual mass distribution function to evaluate the influence of particle density in KBM. The diffusion-based method (DBM) is another smoothing technique based on solving diffusion equations for certain field quantities, which exhibits a similar performance to KBM~\cite{sun2015diffusion}. Capecelatro and Desjardins~\cite{capecelatro2013euler} introduced a KBM-DBM combined method to transfer the particle data onto fluid grids through a kernel-based mollification operation and then dispersed the grid-based data by solving a diffusion equation. Sun and Xiao~\cite{sun2015diffusion,sun2015diffusionapp} proposed a modified version of DBM that replaces the first-step mollification operation~\cite{capecelatro2013euler} with an initial PCM field and then employs an analytical diffusion solution on the volume fraction field with comparable effects of the Gaussian-based KBM. 
Recently, Eshraghi et al.~\cite{eshraghi2023coarse} pointed out the limitations of conventional KBM and DBM when the grid is relatively coarse, and proposed a dynamic KBM inspired by the force interpolation algorithm in the immersed boundary method. This method can adaptively modify the property of kernel function based on the grid size and provide more robust and accurate results under various grid configurations.

Despite many considerations that have been proposed to improve the performance of volume-averaged CFD-DEM coupling, these operations commonly rely on the grid-based direct mapping between discrete particles (their centroids or residing points) and fluid grids (their centers); see Fig.~\ref{fig_compareAllmethod}. Such particle-to-grid-center mapping inevitably causes grid size dependence on the coupling accuracy (specifically in fine grids for local coupling~\cite{wang2019semi} and coarse grids for non-local coupling~\cite{eshraghi2023coarse}). More significantly, it fails to detect subtle yet crucial volumetric changes in very dense granular systems, for which the induced pore fluid pressure plays an important role~\cite{rondon2011granular,pailha2008initiation,pailha2009two}. As a result, common volume-averaged CFD-DEM approaches face difficulties when handling high solid concentrations~\cite{cheng2021resolved}. 

In this work, we address these limitations by proposing a two-step mapping algorithm that uses an intermediate coarse-graining layer between CFD and DEM. In the first step, information carried by DEM particles is converted into smooth, coarse-grained flow fields via Fibonacci sequence point clouds; this discrete-to-continuum step is achieved independently of the resolution of CFD grids. Then, the coarse-grained continuum fields are coupled with the fluid fields through point-based projection to achieve accurate computation of fluid-particle interactions. The proposed method is first verified through three numerical tests with increasing complexity, and then validated against three public data involving both physical experiments and resolved numerical simulations. Compared to existing methods, our CFD-DEM algorithm essentially reduces the influence of the particle-to-grid size ratio on the coupling accuracy and demonstrates significant improvements in simulating dilute to dense particle-fluid systems.

\section{Mathematical description of CFD-DEM framework} \label{sec:method}

We start with a general mathematical description of unresolved and semi-resolved CFD-DEM frameworks, including the volume-averaged N-S equations for the fluid phase, Newton's laws of motion for DEM particles, and the two-way coupling between fluid and particle phases. Note that various sets of CFD-DEM governing equations are available and we generally follow the so-called model A or set II formulation, as reviewed by Zhou et al.~\cite{zhou2010discrete}.

\subsection{Governing equations for fluid phase} \label{sec:method_fluid}

For the fluid phase, a volume filtering operator is applied to the N-S equations to derive the volume-averaged N-S equations~\cite{anderson1967fluid,capecelatro2013euler,zhou2010discrete},
\begin{equation}\label{equ continuous equ}
\frac{\partial \left( \varepsilon _f{\rho _f} \right)}{\partial t}+\nabla \cdot \left( \rho _f\varepsilon _f\boldsymbol{u} \right) =0
\end{equation}
\begin{equation}\label{equ momentum equ}
\frac{\partial \left( \varepsilon _f\rho _f\boldsymbol{u} \right)}{\partial t}+\nabla \cdot \left( \rho _f\varepsilon _f\boldsymbol{u} \otimes \boldsymbol{u} \right) =-\varepsilon _f\nabla p+\varepsilon _f\nabla \cdot \boldsymbol{\tau }-\boldsymbol{F}_{p-f}+\rho _f\varepsilon _f\boldsymbol{g}
\end{equation}
where $\rho _f$ and $\varepsilon _f$ are the fluid density and volume fraction, respectively, $\boldsymbol{u}$ the fluid velocity, $p$ the fluid pressure, $\boldsymbol{\tau}$ the viscous stress tensor, $\boldsymbol{g}$ the gravitational acceleration, and $\boldsymbol{F}_{p-f}$ the particle-fluid interaction force per unit volume (given below). For Newtonian fluids, 
\begin{equation}\label{equ tao}
\boldsymbol{\tau }=\mu _f \left[ \left( \nabla \boldsymbol{u} \right) +\left( \nabla \boldsymbol{u} \right) ^T-2/3\left( \nabla \cdot \boldsymbol{u} \right) \boldsymbol{I} \right] 
\end{equation}
where $\boldsymbol{I}$ is the identity tensor and $\mu_f$ is the dynamic viscosity. In this study, the fluid stress tensor is modeled under the assumption of laminar flow and no turbulence closure is employed.

\subsection{Governing equations for particle phase} \label{sec:method_particle}

The translational and rotational motions of particles are solved by Newtown's laws of motion based on the soft ball model in DEM~\cite{cundall1979discrete}. For particle $i$, 
\begin{equation}\label{equ translational motion}
m_i\frac{\textrm{d}\boldsymbol{v}_i}{\textrm{d} t}=m_i\boldsymbol{g}+\sum_{j=1}^{n_c}{\boldsymbol{f}_{c,ij}}+\boldsymbol{f}_{f-p,i}
\end{equation}
\begin{equation}\label{equ rotational motion}
I_i\frac{\textrm{d}\boldsymbol{\omega }_i}{\textrm{d}t}=\sum_{j=1}^{n_c}{\boldsymbol{T}_{ij}}
\end{equation}
where $n_c$, ${m}_i$, $I_i$, $\boldsymbol{v}_i$, $\boldsymbol{\omega}_i$ are the number of contacts, mass, moment of inertia, linear velocity, and angular velocity, respectively, while $\boldsymbol{f}_{c,ij}$ and ${\boldsymbol{T}_{ij}}$ are the contact force and moment exerted by a neighbor particle $j$, respectively. The force $\boldsymbol{f}_{c,ij}$ is composed of a normal and a tangential components, with the latter limited by the Coulomb friction criterion $|\boldsymbol{f}_{c,ij}^t| \leq \mu_p |\boldsymbol{f}_{c,ij}^n|$, where $\mu_p$ is the coefficient of inter-particle friction.
Torques arising from fluid–particle interactions are neglected, as they are expected to have a negligible impact on the present study involving spherical particles. Interested readers are referred to the work by Eshraghi et al.~\cite{eshraghi2023coarse}.
The interaction force exerted by the fluid, $\boldsymbol{f}_{f-p,i}$, is given next.

\subsection{Coupling between fluid and particle phases} \label{sec:method_forces}

The coupling between CFD and DEM involves the calculation of volume fraction $\varepsilon_f$ and the interaction terms $\boldsymbol{F}_{p-f}$ and $\boldsymbol{f}_{f-p}$ in the governing equations. Typically, closure models are required to express the interaction terms as various fluid-particle interaction forces. For each particle,
\begin{equation}\label{equ hydraulic force on particle}
\boldsymbol{f}_{f-p}=\boldsymbol{f}_{\nabla p}+\boldsymbol{f}_{\nabla \cdot \tau}+\boldsymbol{f}_{drag}+\boldsymbol{f}_{vm}+\boldsymbol{f}_{Basset}+\boldsymbol{f}_{Saff}+\boldsymbol{f}_{Mag}
\end{equation}
where the right-hand side terms are the pressure-gradient force (which includes the buoyancy force), viscosity force, drag force, virtual mass force, Basset force, Saffman lift force and Magnus force, respectively. For the fluid phase,
\begin{equation}\label{equ interaction force}
\boldsymbol{F}_{p-f}=\frac{1}{\Delta V}\sum_{i=1}^{n_p}{\left( \boldsymbol{f}_{drag,i}+\boldsymbol{f}_{vm,i}+\boldsymbol{f}_{Basset,i}+\boldsymbol{f}_{Saff,i}+\boldsymbol{f}_{Mag,i} \right)}
\end{equation}
where $ \Delta V $ is a control volume, $n_p$ is the number of particles in the control volume, and $i$ is the index for particles. Note that for kernel-based CFD-DEM approaches, specific treatments are needed to distribute the particle volume into a number of fluid grids, leading to modification of the $\boldsymbol{F}_{p-f}$ formulation. Comparing to Eq.~\eqref{equ hydraulic force on particle}, $\boldsymbol{f}_{\nabla p}$ and $\boldsymbol{f}_{\nabla \cdot \tau}$ are not explicitly included in Eq.~\eqref{equ interaction force} because they are merged with pressure gradient and viscous stress terms Eq.~\eqref{equ momentum equ}, known as the model A or set II formulation~\cite{zhou2010discrete}. Except for the single-sphere sedimentation test (Section~\ref{validate:sedimentation}) 
, where nearly matched fluid and particle densities are used and the virtual mass force becomes non-negligible, we do not consider the last four force terms in Eqs.~\eqref{equ hydraulic force on particle} and~\eqref{equ interaction force} in the remainder of this paper. Nevertheless, they can be treated similarly to the drag term when needed. For the drag force on particle $i$, we use the Gidaspow model~\cite{gidaspow1994multiphase},
\begin{equation}\label{equ gidaspow drag}
\boldsymbol{f}_{drag,i}=\frac{1}{6}\pi d^3\frac{\beta}{1-\varepsilon _f}\left( \boldsymbol{u}-\boldsymbol{v}_i \right) 
\end{equation}
\begin{equation}\label{equ:gidaspow-drag-coe}
\beta =
\begin{cases}
\displaystyle
150 \dfrac{(1 - \varepsilon_f)^2}{\varepsilon_f} \dfrac{\mu_f}{d^2}
+ 1.75 (1 - \varepsilon_f) \dfrac{\rho_f}{d}
\left| \boldsymbol{u} - \boldsymbol{v}_i \right|, 
& \varepsilon_f \leqslant 0.8 \\[8pt]
\displaystyle
\dfrac{3}{4} C_d \dfrac{\rho_f \varepsilon_f (1 - \varepsilon_f)}{d}
\left| \boldsymbol{u} - \boldsymbol{v}_i \right| \varepsilon_f^{-2.65}, 
& \varepsilon_f > 0.8
\end{cases}
\end{equation}

\begin{equation}\label{equ:Cd}
C_d =
\begin{cases}
\displaystyle
\dfrac{24}{\mathrm{Re}_p} \left( 1 + 0.15\, \mathrm{Re}_p^{0.687} \right), 
& \mathrm{Re}_p < 1000 \\[6pt]
0.44, 
& \mathrm{Re}_p \geqslant 1000
\end{cases}
\end{equation}
where $
\text{Re}_p=\varepsilon _f\rho _fd\left| \boldsymbol{u}-\boldsymbol{v}_i \right|/\mu _f
$ is the particle Reynolds number, $d$ is the particle diameter, and $C_d$ is the drag coefficient. This model includes the Ergun equation~\cite{ergun1952fluid} for dense conditions ($\varepsilon _f \leqslant 0.8$) and Wen \& Yu model~\cite{wen1966mechanics} for dilute conditions ($\varepsilon _f > 0.8$), whereas the $\varepsilon _f^{-2.65}$ term accounts for the group effect on the drag coefficient of a single sphere. Finally, although different drag force models exist~\cite{liu2022general}, using other drag models does not affect the main conclusions of this paper.

\section{Two-step mapping strategy for CFD-DEM via coarse graining}\label{sec:methodCG}

\begin{figure}[htbp]
        \centering
        \includegraphics[width=\textwidth,height=9cm, keepaspectratio=true]{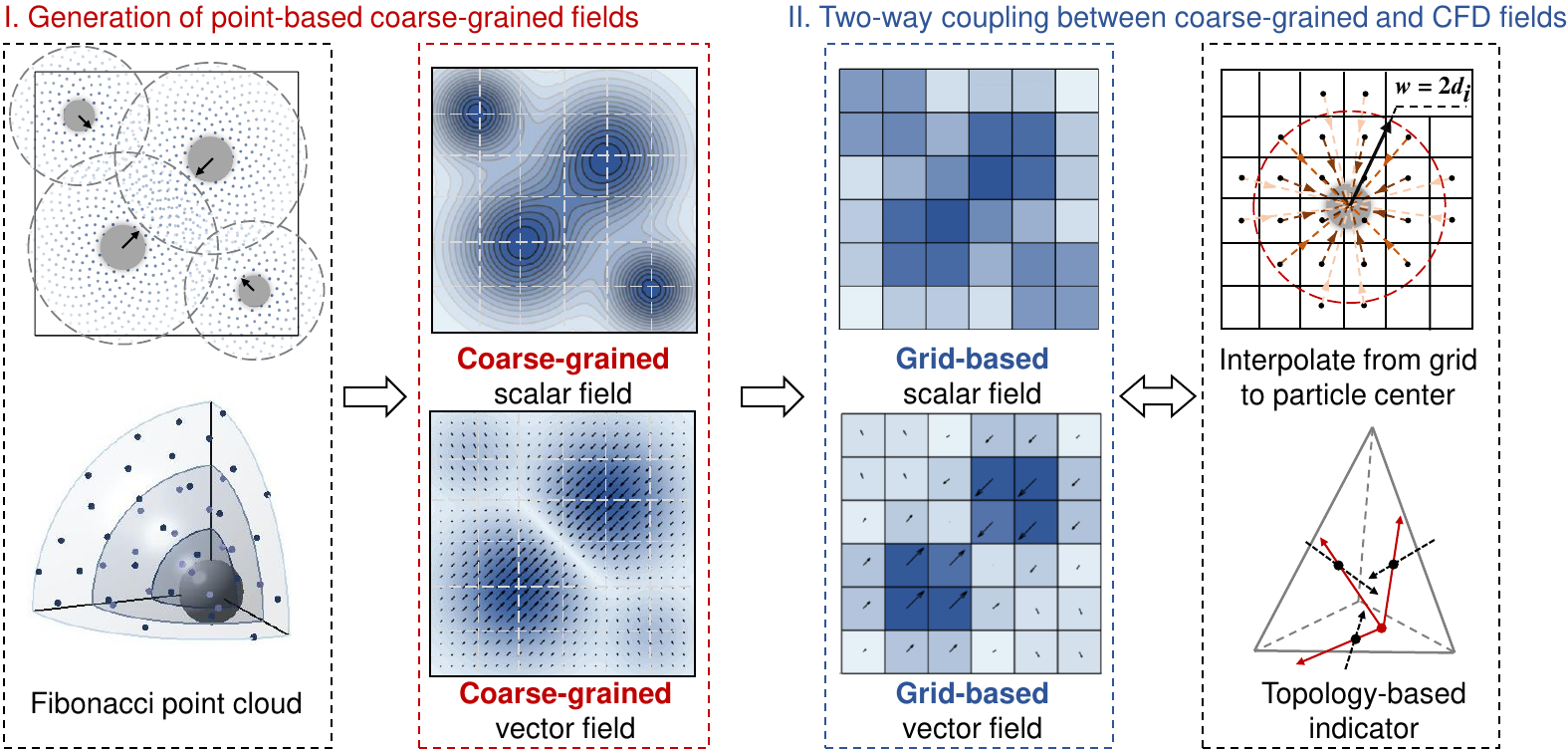}
     \caption{\label{fig CG overview} Overview of the two-step coarse-graining CFD-DEM method, which consists of a point-based dispersion for generating the coarse-grained fields and a point-based projection for mapping information between CFD and DEM. The kernel width varies with the particle, as it is set to be two times the particle diameter. 
     }
\end{figure}

As discussed in Section~\ref{sec:intro}, a common challenge in unresolved and semi-resolved CFD-DEM is to map particle information onto the fluid grids without a strong dependence on the particle and grid sizes. We address this issue by proposing a two-step mapping strategy that uses a novel intermediate coarse graining step to achieve accurate and grid-independent coupling (see Fig.~\ref{fig CG overview}). We first briefly introduce the concept of coarse graining as a natural smoothing technique for discrete particles (Section \ref{sec:methodCG_intro}), then explain how point clouds are employed to construct coarse-grained fields from the particle-based data (Section~\ref{sec:methodCG_point}) and, finally, describe the two-way coupling between coarse-grained and CFD fields to complete the CFD-DEM coupling (Section \ref{sec:methodCG_coupling}).

\subsection{Coarse graining: from particle information to continuum fields}\label{sec:methodCG_intro}

In statistical mechanics, coarse graining is a method to convert discrete particle data (position, velocity, forces) into continuum fields, which has the advantages of automatically satisfying the conservation equations of continuum mechanics and being generically smooth for even a single-configuration of particles \cite{goldhirsch2010stress,weinhart2012discrete}. It has been widely used in DEM and molecular dynamics for smoothening the data or computing the continuum fields, and was recently used for data mapping in a concurrent continuum-discrete coupling method \cite{cheng2023concurrent}. It has not yet been explicitly employed to convert particle-level data into continuum fields within the CFD-DEM coupling procedure, though the non-local coupling strategies reviewed in Fig.~\ref{fig_compareAllmethod} share some features similar to coarse graining.

For a set of particles centered at position $\boldsymbol{x}_i$, the microscopic mass density field $\rho^{mic}$ at a point $\boldsymbol{x}$ and time $t$ can be defined by
\begin{equation}\label{eq:rho_mic}
    \rho ^{mic}\left( \boldsymbol{x},t \right) =\sum_{i=1}^{N_p}{m_i\delta \left( \boldsymbol{x}-\boldsymbol{x}_i\left( t \right) \right)}
\end{equation}
where $N_p$ is the number of particles in the computational domain, $\delta(\cdot)$ is the Dirac delta function, and $m_i$ is the mass of particle $i$. This definition satisfies that the integral of the mass density over a given volume equals the total mass, but it represents a singular (i.e., point-like) entity. The coarse-grained mass density field $\rho$ for the granular phase can then be extracted by convoluting $ \rho ^{mic}\left( \boldsymbol{x},t \right)$ with a kernel function $\mathcal{W}$~\cite{goldhirsch2010stress}, yielding
\begin{equation}\label{equ:CG kernelbased}
\rho \left( \boldsymbol{x}, t \right) =\rho ^{mic}\left( \boldsymbol{x}, t \right) *\mathcal{W}=\sum_{i=1}^{N_p}{m_i\int_{\mathbb{R}^3}{\delta \left( \boldsymbol{x'}-\boldsymbol{x}_i\left( t \right) \right)}\mathcal{W}\left( \boldsymbol{x}-\boldsymbol{x'} \right)}d\boldsymbol{x'}
\end{equation}
where $*$ denotes a convolution and $\mathcal{W}$ is a kernel function or a coarse graining function, which has an integral of unity over the domain (hence $\mathcal{W}$ has the units of inverse volume) and a coarse-graining width $w$. There are various choices for $\mathcal{W}$~\cite{weinhart2016influence} and we use a truncated Gaussian distribution (given below).

Following this procedure, the coarse-grained fields of volume fraction $\phi$, momentum density $\boldsymbol{J}$, and velocity $\boldsymbol{v}$ for the granular phase are given by
\begin{equation}\label{equ CG volume}
\phi \left( \boldsymbol{x},t \right) =\sum_{i=1}^{N_p}{\frac{m_i}{\rho_{p,i}}\mathcal{W}\left( \boldsymbol{x}-\boldsymbol{x}_i\left( t \right) \right)}
\end{equation}
\begin{equation}\label{equ CG momentum}
\boldsymbol{J}\left( \boldsymbol{x},t \right) =\sum_{i=1}^{N_p}{m_i\boldsymbol{v}_i\mathcal{W}\left( \boldsymbol{x}-\boldsymbol{x}_i\left( t \right) \right)}
\end{equation}
\begin{equation}\label{equ CG velocity}
\boldsymbol{v}\left( \boldsymbol{x},t \right) =\frac{\boldsymbol{J}\left( \boldsymbol{x},t \right)}{\rho \left( \boldsymbol{x},t \right)}
\end{equation}
where $\rho_{p,i}$ is the material density of particle $i$. Note that $\boldsymbol{v}$ is a derived quantity in coarse graining.

Other coarse-grained fields, such as the contact stress tensor and granular temperature, can be similarly extracted considering the momentum balance \cite{goldhirsch2010stress,weinhart2012discrete}. While they are omitted here, future work considering granular stress or temperature in fluid-particle interactions may include these coarse-grained fields in the coupling process.

\subsection{Generating coarse-grained fields via point cloud-based dispersion} \label{sec:methodCG_point}

In principle, coarse-grained fields can be evaluated at an arbitrary resolution (i.e., at any location $\boldsymbol{x}$). In practice, however, a set of evaluation points need to be defined in association with the grid resolution (which is somewhat similar to the kernel-based methods in Fig.~\ref{fig_compareAllmethod})  or the particle centers. Here, to avoid grid dependence in the coarse graining process, we use particle-based point clouds to serve as the evaluation points, which essentially disperses the particle information over a range set by the kernel function without considering the fluid grids. 

\begin{figure}[htbp]
        \centering
        \includegraphics[width=\textwidth,height=5cm, keepaspectratio=true]{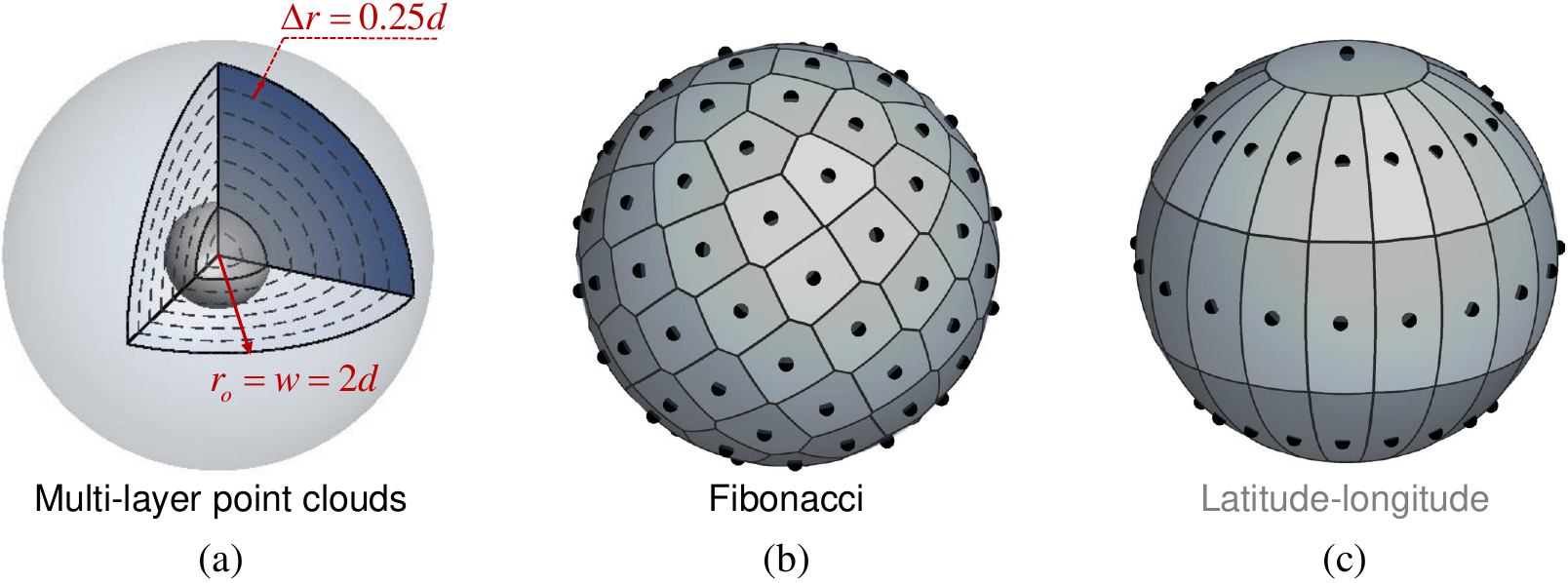}
     \caption{\label{fig Fibonacci_comp} Geometry of the multi-layer point cloud associated with one particle.  (a) Radii and spacing of the multiple layers. (b) and (c) compare Voronoi tessellations constructed from Fibonacci and latitude–longitude point clouds; the former is clearly more uniform and isotropic.
     }
\end{figure}

The point clouds we employ consist of multiple equally spaced spherical layers (Fig.~\ref{fig Fibonacci_comp}). Let $N_l$ be the number of points on the $l$-th layer. To evenly distribute the evaluation points on the spherical surface, we use the Fibonacci lattice~\cite{wang2024computational,chukkapalli1999scheme,baselga2018fibonacci} to determine the Cartesian coordinates of the $n$-th point (on a unit sphere centered at the origin),
\begin{equation} \label{equ fibonacci z}
    z_n=\frac{2n-1}{N_l}-1
\end{equation}
\begin{equation} \label{equ fibonacci x}
    x_n=\sqrt{1-z_{n}^{2}}\cdot \cos \left( 2\pi n\xi \right) 
\end{equation}
\begin{equation} \label{equ fibonacci y}
    y_n=\sqrt{1-z_{n}^{2}}\cdot \sin \left( 2\pi n\xi \right) 
\end{equation}
where $\xi =(\sqrt{5}-1)/2$ is Fibonacci's golden ratio, 
$N_l=N_o \left( r_l/r_o \right) ^2$, $r_l$ is the radius of the $l$-th spherical layer and the subscript $o$ denotes the outermost layer. The formula for $N_l$ is determined as each point in the three-dimensional Fibonacci lattice occupies an area corresponding to its equal-area Voronoi grid on the spherical surface~\cite{swinbank2006fibonacci}; see Fig.~\ref{fig Fibonacci_comp} for an illustration of the Voronoi tessellation. In our implementation, we set $r_o=w=2d$ (tested in Section~\ref{sec:benchmarks_static}), in line with previous studies~\cite{zhu2022semi,wang2021determination,eshraghi2023coarse}, where values from $2d_i$ to $3d_i$ are typically used.
The space between two adjacent layers is $\Delta r=0.25d$ and the number of points on the outermost layer is $N_o = 96$, leading to nearly $300$ points for each particle. As demonstrated in Figs.~\ref{fig Fibonacci_comp}b and c, the advantage of the Fibonacci lattice is that the latitude coordinates of each point, $z_n$, form an arithmetic sequence allowing each point to be located on an independent horizontal level~\cite{saff1997distributing}. This makes the Fibonacci point clouds virtually more uniform and isotropic than the more straightforward latitude-longitude point clouds~\cite{gonzalez2010measurement}.

As the point clouds are defined, the coarse graining function $\mathcal{W}$ for particle $i$ can now be evaluated at its associated point $k$, whose position is $\boldsymbol{x}_k$,
\begin{equation}\label{eqn:gaussiankernel}
\mathcal{W}_{ik}=C\exp \left( -\frac{\left| \boldsymbol{x}_i-\boldsymbol{x}_k \right|^2}{2w^2} \right) 
\end{equation}
which is a truncated Gaussian kernel with width $w$ and a re-normalization coefficient
\begin{equation}\label{eqn:gaussiankernel_C}
C= \left[ \sum_{k=1}^{N_{pt}}\exp \left( -\frac{\left| \boldsymbol{x}_i-\boldsymbol{x}_k \right|^2}{2w^2} \right) \right]^{-1}
\end{equation}
where $N_{pt}$ is the number of points associated with particle $i$. Using this expression, the (partial) coarse-grained mass density, volume fraction, and momentum density, contributed by particle $i$ at point $k$, are given respectively by
\begin{equation}
\rho_{ik}=m_i\mathcal{W}_{ik}
\end{equation}
\begin{equation}
\phi _{ik}=\frac{m_i}{\rho_{p,i}}\mathcal{W}_{ik}
\end{equation}
\begin{equation}\label{equ:point-based momentum}
\boldsymbol{J}_{ik}=m_i\boldsymbol{v}_i\mathcal{W}_{ik}
\end{equation}
This point cloud dispersion process is independent of the fluid grids and is repeated for all particles. The resulting point-based coarse-grained fields are coupled with the fluid grids in the next step.



\subsection{Coupling between coarse-grained fields and fluid grids via point-based projection}\label{sec:methodCG_coupling}

\begin{figure}[htbp]
        \centering
        \includegraphics[width=\textwidth,height=9cm, keepaspectratio=true]{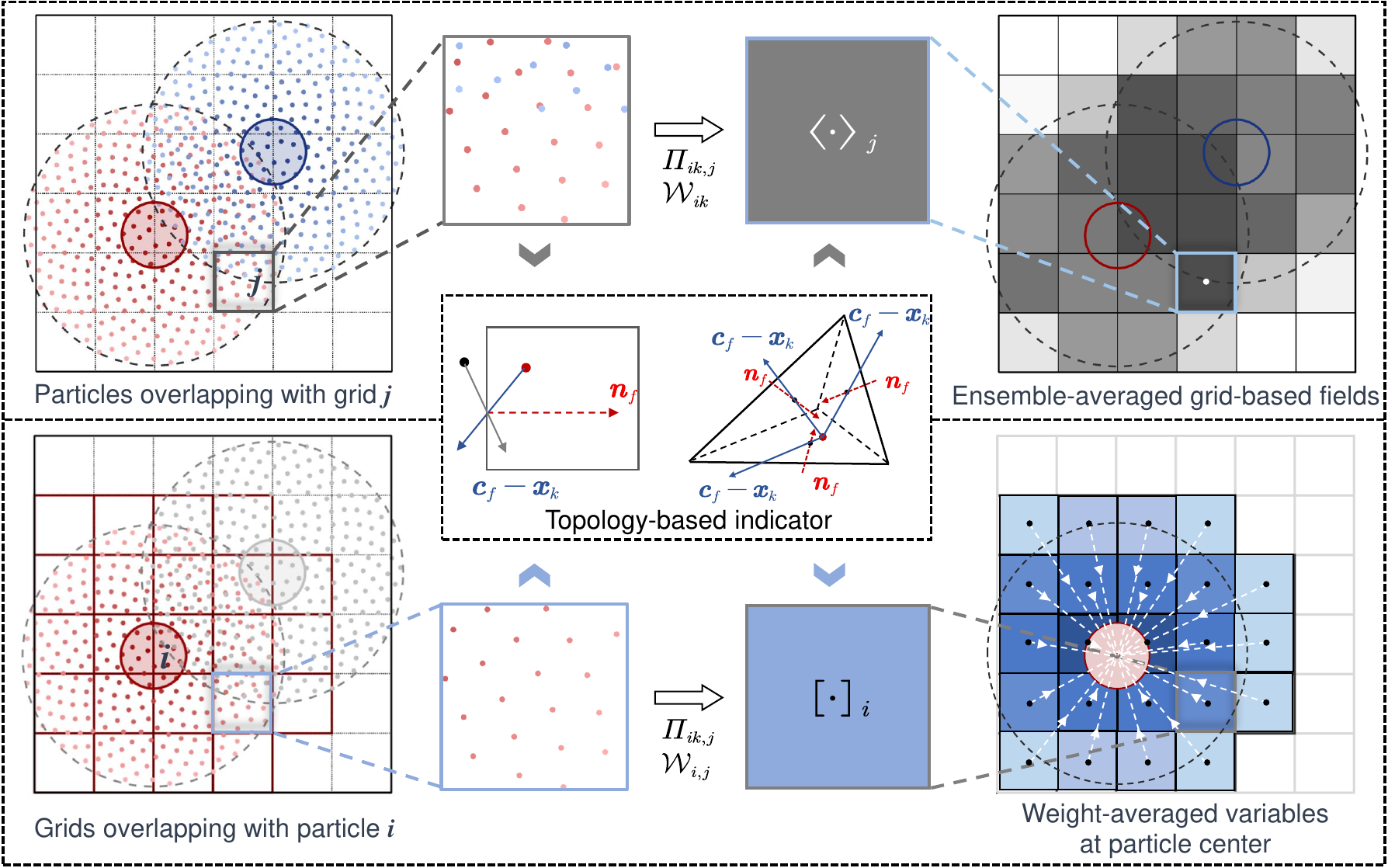}
     \caption{\label{fig CG couple} Coupling between point-based coarse-grained fields and fluid grid variables. The upper panel corresponding to Eqs.~\eqref{eq:gridbased volume-fraction} and~\eqref{eq:gridbased particle-velocity} shows the calculation of the grid-averaged quantities for closing the fluid governing equations. The lower panel corresponding to Eqs.~\eqref{equ volume-averaged weight},~\eqref{equ particle-based void},~\eqref{equ particle-based fluid velocity}, and~\eqref{equ drag force on particle} shows the calculation of the weight-averaged variables at the particle center for the drag force calculation. The central dashed box illustrates the topology-based indicator, i.e., Eq.~\eqref{equ determine point-in-grid}.
     }
\end{figure}

In conventional CFD-DEM methods, coupling is often achieved between a fluid grid and the particles (or parts of their expansion domains) contained in the grid. Here, by dispersing the particle-based discrete data into point cloud-based continuum fields, it becomes convenient to perform two-way coupling between a fluid grid and all its enclosing points (representing parts of different particles inside the grid), as illustrated in Fig.~\ref{fig CG couple}. To do so, we define a topology-based indicator, $\varPi_{ik,j}$, which determines whether point $k$ associated with particle $i$ lies in the domain of fluid grid $j$,
\begin{equation}\label{equ determine point-in-grid}
\varPi _{ik,j}=
    \begin{cases}
    	1&		\boldsymbol{n}_f\cdot \left( \boldsymbol{c}_f-\boldsymbol{x}_k \right) \leqslant 0, \forall f\in \partial R_j\\
    	0&		\textrm{otherwise}\\
    \end{cases}
\end{equation}
where $\partial R_{j}$ represents all boundary faces enclosing grid $j$, $\boldsymbol{n}_f$ is the normal vector to grid face $f$ (oriented from the neighbour to the owner grid), and $\boldsymbol{c}_f$ is the center position vector of grid face $f$. For structured grids, the indicator can be simplified to a straightforward coordinate retrieval process using a look-up table, significantly improving the computational efficiency (Section~\ref{sec: efficiency}).

It is now possible to compute grid-averaged quantities for closing the fluid governing equations, including the fluid volume fraction $\left<\varepsilon\right>_j$ and particle velocity $\left<\boldsymbol{v}\right>_j$, 
\begin{equation}\label{eq:gridbased volume-fraction}
\left< \varepsilon \right>_j = 1-\frac{1}{V_j}{\sum_{i=1}^{N_p}}{\sum_{k=1}^{N_{pt}}} \left( \phi_{ik} \varPi_{ik,j} \right)
\end{equation}
\begin{equation}\label{eq:gridbased particle-velocity}
\left< \boldsymbol{v} \right> _j=\frac{\sum_{i=1}^{N_p}{\sum_{k=1}^{N_{pt}} \left( \boldsymbol{J}_{ik} \varPi_{ik,j}  \right) }}{\sum_{i=1}^{N_p}{\sum_{k=1}^{N_{pt}}\left(  \rho_{ik} \varPi_{ik,j} \right) }}
\end{equation}
where $\left< \cdot \right>_j$ represents an ensemble average over grid $j$, $V_j$ is the grid volume, and $N_p$ is the number of particles in the simulation. Note that $\left< \varepsilon \right>_j$ is used in the fluid governing equations as the porosity field, whereas $\left<\boldsymbol{v}\right>_j$ is used as part of the semi-implicit momentum exchange term. It is necessary to convert discrete particle velocities into a continuous field when using a semi-implicit scheme (see~\ref{app:stability} for details), but this is often overlooked in the volume-averaged CFD-DEM frameworks.

To compute the interaction forces acting on the particles, various fluid grid-based quantities need to be projected onto the particle center, which again relies on the weights $\mathcal{W}_{ik}$ carried by the point clouds (see Fig.~\ref{fig CG couple}, lower panel). Consider particle $i$, which has a dispersion domain overlaps with $N_g$ fluid grids. The total normalized weights assigned to grid $j$ is denoted by
\begin{equation}\label{equ volume-averaged weight}
\mathcal{W}_{i,j} = \sum_{k=1}^{N_{pt}}{ \left( \mathcal{W}_{ik} \varPi_{ik,j} \right) }
\end{equation}
which is used to compute the following weighted-averaged variables at the particle center,
\begin{equation}\label{equ particle-based void}
\left[ \varepsilon \right]_i=\sum_{j=1}^{N_g}{ \left( \left< \varepsilon \right>_j \mathcal{W}_{i,j} \right) }
\end{equation}
\begin{equation}\label{equ particle-based fluid velocity}
\left[ \boldsymbol{u} \right]_i=\sum_{j=1}^{N_g}{ \left( \boldsymbol{u}_j \mathcal{W}_{i,j} \right) }
\end{equation}
where $\left[ \cdot \right]_i$ indicates weighted interpolation of a grid-based variable at the center of particle $i$. Note that $\boldsymbol{u}_j$ is the fluid velocity of grid $j$, not to be confused with $\boldsymbol{v}_i$, which is the velocity of particle $i$. These variables are substituted into the Gidaspow drag model, Eq.~\eqref{equ gidaspow drag}, to compute the drag force acting on particle $i$,
\begin{equation}\label{equ drag force on particle}
\boldsymbol{f}_{drag,i}=\frac{1}{6}\pi d_i^3\frac{\beta}{1-\left[ \varepsilon \right]_i}\left( \left[ \boldsymbol{u} \right] _i-\boldsymbol{v}_i \right) 
\end{equation}
The overall interaction force exerting on a particle is $\boldsymbol{f}_{f-p}=\boldsymbol{f}_{\nabla p}+\boldsymbol{f}_{\nabla \cdot \tau}+\boldsymbol{f}_{drag}$, as detailed in Section~\ref{sec:method_forces}, but the weighted interpolation is only applied to $\boldsymbol{f}_{drag}$ in the present work. Other forces (e.g., $\boldsymbol{f}_{\nabla p}$ and $\boldsymbol{f}_{\nabla \cdot \tau}$) can be treated similarly when necessary.

Following the model A formulation (Section~\ref{sec:method}), the particle-to-fluid interaction term in the fluid governing equations for grid $j$, $\left< \boldsymbol{F}_{p-f} \right>_j$, is essentially a volume-averaged drag force, 
\begin{equation}\label{equ force to fluid}
\left< \boldsymbol{F}_{p-f} \right>_j=\frac{1}{V_j}\sum_{i=1}^{N_p}{\left( \boldsymbol{f}_{drag,i} \mathcal{W}_{i,j} \right) }
\end{equation}
where $N_p$ is the number of particles whose expansion domains overlap with grid $j$. 

\begin{figure}[htbp]
        \centering
        \includegraphics[width=\textwidth,height=10cm, keepaspectratio=true]{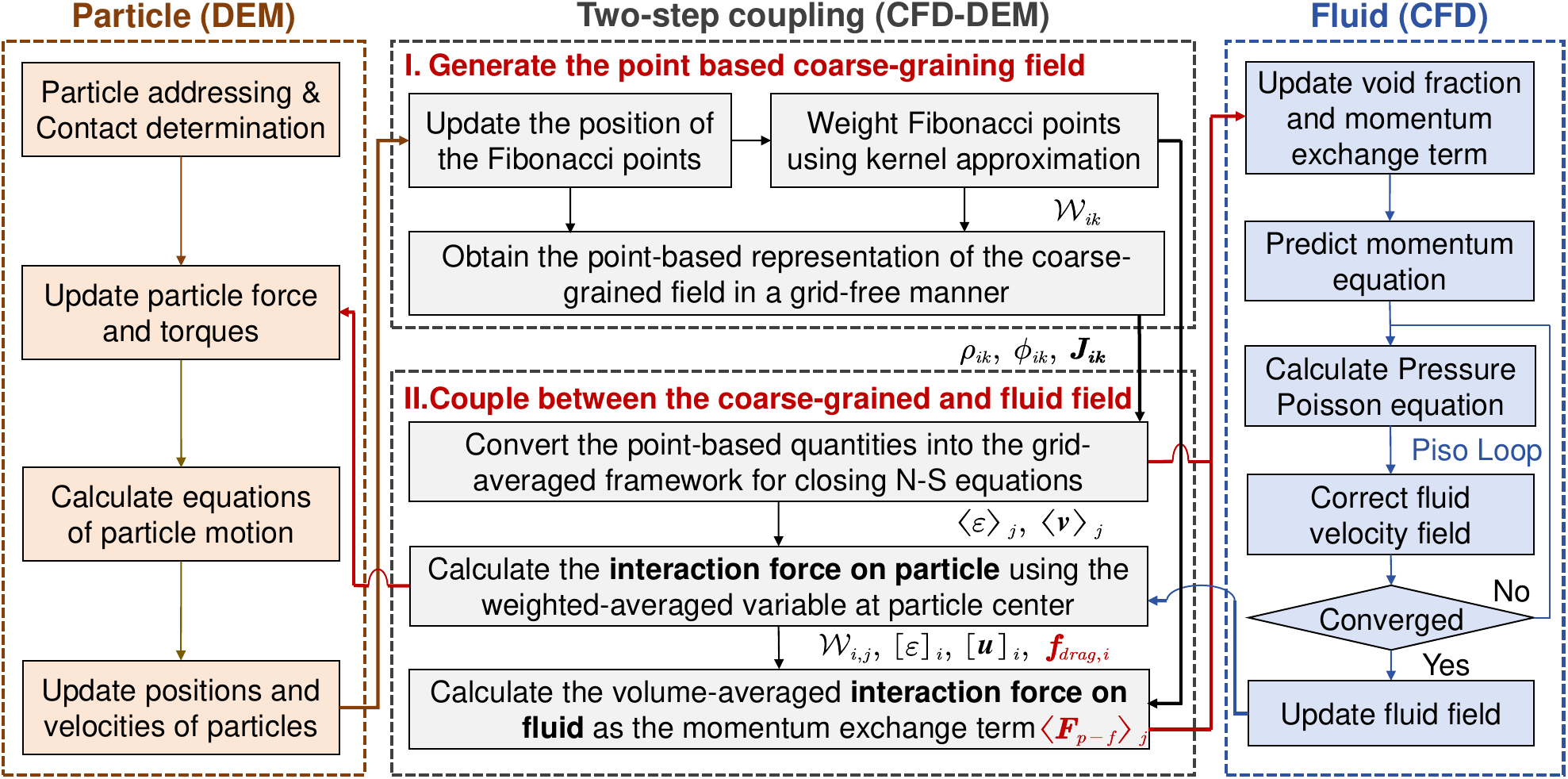}
     \caption{\label{fig flowchat} Flow chart of the proposed CFD-DEM method. Coarse graining is used as an intermediate step for two-way coupling.}
\end{figure}

In summary, using the coarse-graining functions (weights) evaluated at the point clouds, we compute fluid grid-averaged quantities (volume fraction $\left< \varepsilon \right>_j$ and particle-to-fluid interaction term $\left< \boldsymbol{F}_{p-f} \right>_j$) to close the CFD equations and the particle-based interaction forces $\boldsymbol{f}_{f-p,i}$ to close the DEM equations.
Fig.~\ref{fig flowchat} presents a flow chart of the proposed CFD-DEM method. The numerical scheme is implemented within the open-source library CFDEM framework, combining OpenFOAM and LIGGGHTS for CFD and DEM parts, respectively. The extended pressure implicit with splitting of operators (PISO) algorithm for the volume-averaged N-S equations is adopted to solve the fluid governing equations. The second-order central difference scheme (Gauss linear) and the implicit first-order scheme (Euler) are adopted for spatial and temporal discretization, respectively.

\section{Benchmark tests of numerical accuracy and stability} \label{sec:benchmarks}
In this section, we show improved accuracy and stability of the proposed method through three carefully designed numerical tests with increasing complexities. While the proposed method is similar to a standard KBM in that a kernel function (coarse-graining function) is used for spreading the discrete data, it makes use of coarse graining and Fibonacci point clouds to avoid grid dependence and achieve smoothened two-way mapping between the particles and fluid grids. We compare the performance of the proposed method with a standard KBM in all three tests.

\subsection{A static particle on one-dimensional grids} \label{sec:benchmarks_static}

\begin{figure}[htbp]
        \centering
        \includegraphics[width=\textwidth,height=12cm, keepaspectratio=true]{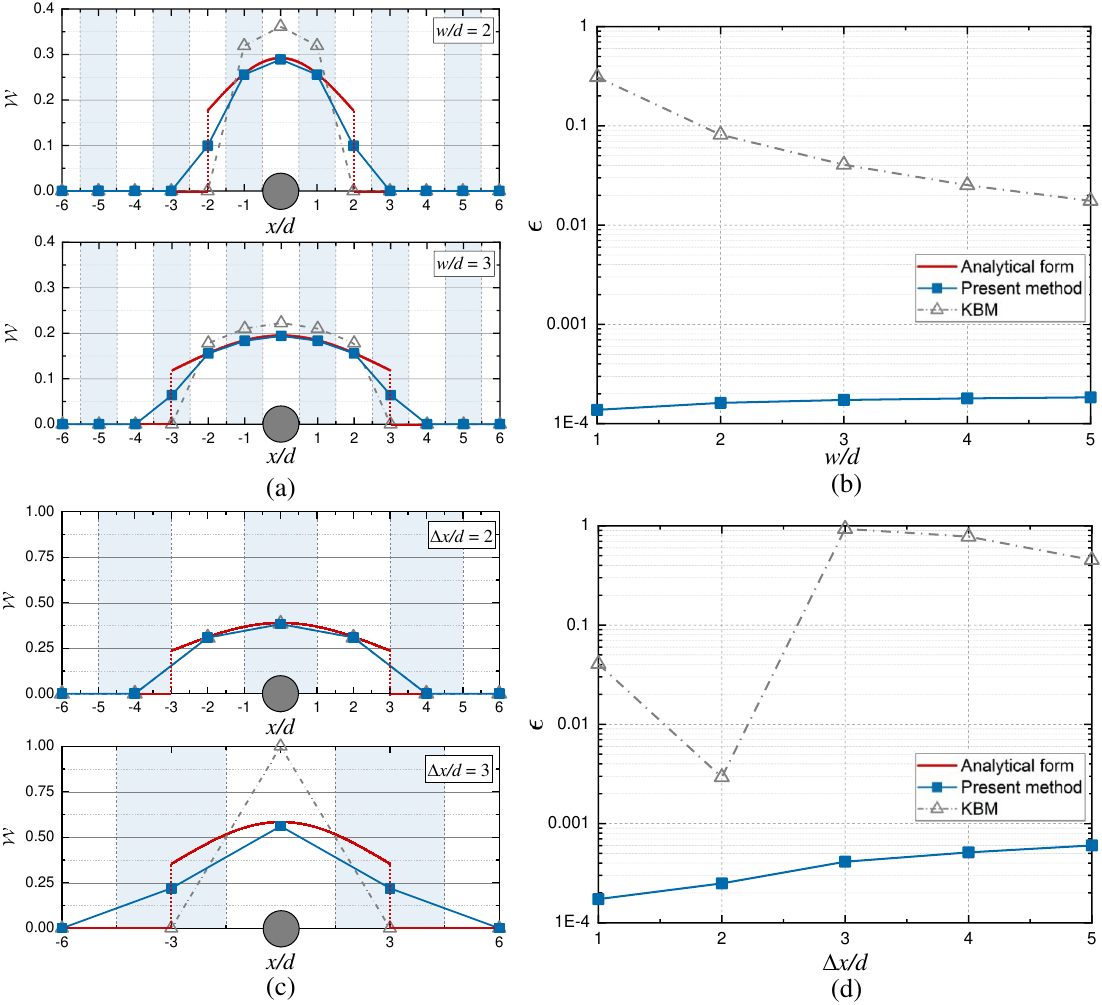}
     \caption{\label{fig:benchmarks_1d} Weight allocation test for a static particle located on one-dimensional grids, comparing the proposed method and KBM. (a) Weight distribution for varied kernel width, $w/d = 2$ and $3$. (b) Errors relative to the analytical form for $w/d = 1 \sim 5$. (c) Weight distribution for varied grid size, $\Delta x/d=2$ and $3$. (d) Error relative to the analytical form for $\Delta x/d=1\sim5$. The grids are indicated by alternating shading and dashed lines for clarity.}
\end{figure}

In coarse graining or kernel-based methods, accurate evaluation of the weights allocated from the particle center to its surrounding grids is crucial to the accuracy of CFD-DEM coupling. To test this fundamental aspect of the proposed method, we design a numerical test where a particle is centered at $x=0$ and its weights $\mathcal{W}(x)$ are evaluated on a series of equally spaced one-dimensional (1D) grids (see Fig.~\ref{fig:benchmarks_1d}). An analytical form of $\mathcal{W}(x)$, which is a truncated Gaussian kernal given by Eq.~\eqref{eqn:gaussiankernel}, is plotted as a red curve for a given kernel width $w$ and grid size $\Delta x$. The proposed two-step mapping method first evaluates $\mathcal{W}$ at a set of Fibonacci points distributed within the expansion domain. Each value is then assigned to the grid in which the point resides. The results are compared with those of the standard KBM, which directly evaluates $\mathcal{W}$ at the grid center.

Fig.~\ref{fig:benchmarks_1d}a shows that, for $\Delta x/d=1$ but varied $w/d$, the proposed method is accurate up to half grid at the edge of the expansion domain, whereas the standard KBM overestimates $\mathcal{W}$ inside the expansion domain, hence a sharper gradient at the edge. This comparison highlights how the proposed method is more accurate and produces smoother results when the grid size is about the particle size ($\Delta x/d=1$). To systematically assess the accuracy of the results for various $\Delta x / d$, we define an overall error $\epsilon$,
\begin{equation}
\epsilon =\sum_{j=1}^{N_g}{\left| \mathcal{W}\left( x_j \right)\Delta{x}-\int_{x_j-\Delta{x}/2}^{x_j+\Delta{x}/2}{\mathcal{W}_\textrm{anal}}\left( x \right) \,dx \right|}
\end{equation}
where $\mathcal{W}\left( {x}_j \right)$ is the weight evaluated by the proposed or conventional method at a grid center $x_j$ and ${\mathcal{W}_\textrm{anal}}\left( x \right)$ indicates the analytical kernel function. As shown in Fig.~\ref{fig:benchmarks_1d}b, the error of KBM is at least two orders of magnitude greater than the proposed method. The error in the KBM primarily arises from two sources. First, the method evaluates the weight distribution only at grid centers, resulting in a limited sampling resolution. Second, truncation error occurs at the boundary of the expansion domain, altering the re-normalization coefficient of the kernel function.  
The accuracy of KBM can be improved by increasing the kernel width $w$, but using too large $w$ will lead to efficiency problems.
It is also noted that the accuracy of the proposed method slightly decreases with increasing grid size. This is because larger grids have relatively fewer representative points per grid, reducing the spatial resolution. 

Likewise, we test the effects of $\Delta x/d$ on weight allocation for fixed $w/d=3$ in Figs.~\ref{fig:benchmarks_1d}c and d. 
We note that the error of KBM is somewhat sensitive to where the particle is located concerning the grid edge and does not change monotonically with the grid resolution (see Fig.~\ref{fig:benchmarks_1d}d). For instance, in a special configuration of $\Delta x/d=2$, $\epsilon$ suddenly drops as the expansion domain precisely covers the boundary grids and the truncation error at the edge of the expansion domain vanishes. Moreover, for a relatively coarse grid configuration, the standard KBM reduces to PCM because one grid can be wider than the expansion domain. The weight calculation in KBM heavily relies on a finite number of grid center points, which leads to divergence relative to the analytical form and a strong dependency on the grid resolution.

In summary, we use a static, 1D configuration to demonstrate the robustness and accuracy of the proposed method in weight allocation for a widely varied $w/d$ and $\Delta x/d$. By contrast, the accuracy of the conventional KBM method depends strongly on $w/d$, $\Delta x/d$, and the relative position between the grid and the particle. This last point indicates that the conventional method may suffer from significant numerical oscillations as the particle is moving across the grids, which is demonstrated in the next test.

\subsection{A moving particle on two-dimensional grids} \label{sec:benchmarks_2d}

\begin{figure}[htbp]
        \centering
        \includegraphics[width=\textwidth,height=8.5cm, keepaspectratio=true]{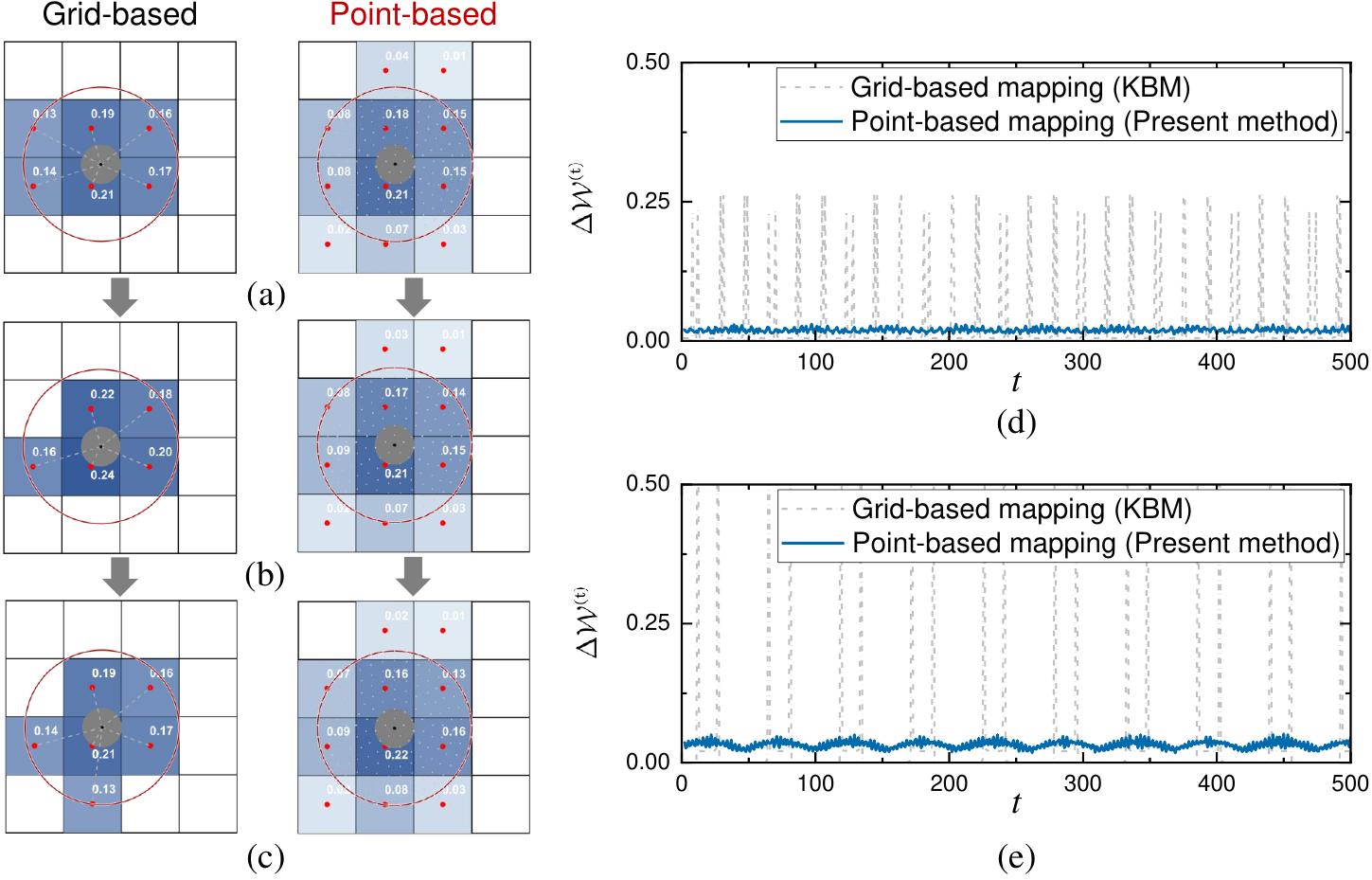}
     \caption{\label{fig:benchmarks_2d} Weight allocation of a falling particle of diameter $d$ (a) at the initial position, (b) after falling by $0.05d$, and (c) after falling by $0.1d$. The temporal variation of the overall weight during the particle falling process is evaluated for (d) fine and (e) coarse grids.}
\end{figure}

We further test the weight allocation of a moving particle on two-dimensional grids (see Fig.~\ref{fig:benchmarks_2d}), which is motivated by two commonly observed issues of conventional CFD-DEM methods, namely, numerical oscillations as a particle is moved across grids and large variations in the fluid fields due to tiny particle displacements. To this end, a particle of diameter $d$ is moved downward successively by a small displacement $0.05d$ every timestep $\Delta t$ and the weight distributions around the particle is evaluated using the proposed method and the standard KBM, respectively.

As shown in Figs.~\ref{fig:benchmarks_2d}a-c, where the weight distributions are indicated by corresponding numbers and shading colors, the standard KBM indeed leads to significant variations in the weight distribution as a small particle displacement occurs, whereas the results of the proposed method is virtually more stable and uniform over time. The major source of error in the KBM is due to the direct mapping from the particle center to the grid center (red dots), which leads to truncation errors at the boundaries of the expansion domain (red circles). By contrast, the proposed method accurately resolves the boundaries of the expansion domain via the point clouds (light points) and capture subtle variations of the weight field.
To better quantify the temporal variations of evaluated weight fields, we calculate the summed weight variations as a function of $t$,
\begin{equation}\label{D_weight}
\Delta \mathcal{W}(t)=\sum_{j=1}^{N_g}{\left| \mathcal{W}_{j}^{\left( t \right)}-\mathcal{W}_{j}^{\left( t-\Delta{t} \right)} \right|}
\end{equation}
where the summation is done over all grids and the superscripts $t$ and $t-\Delta t$ indicate the current and previous timesteps, respectively. 
Figs.~\ref{fig:benchmarks_2d}d and e show $\Delta \mathcal{W}$ as a function of $t$ for two grid resolutions, $\Delta x/d=1.5$ and $3$, respectively. The falling rate is constant at $0.05d/\Delta{t}$. Periodic oscillations of $\Delta \mathcal{W}$ can be observed for the standard KBM, which is exacerbated when coarser grids are used, whereas the proposed method significantly reduces the oscillations for both fine and coarse grids.

This test highlights the good stability, accuracy and grid independence of the proposed two-step mapping method in weight allocation, hence providing a strong foundation for the numerical robustness of our method in the following more complicated test cases.

\subsection{Pressure drop of fluid flow through a dense granular bed}

\begin{figure}[htbp]
        \centering
        \includegraphics[width=\textwidth,height=7cm, keepaspectratio=true]{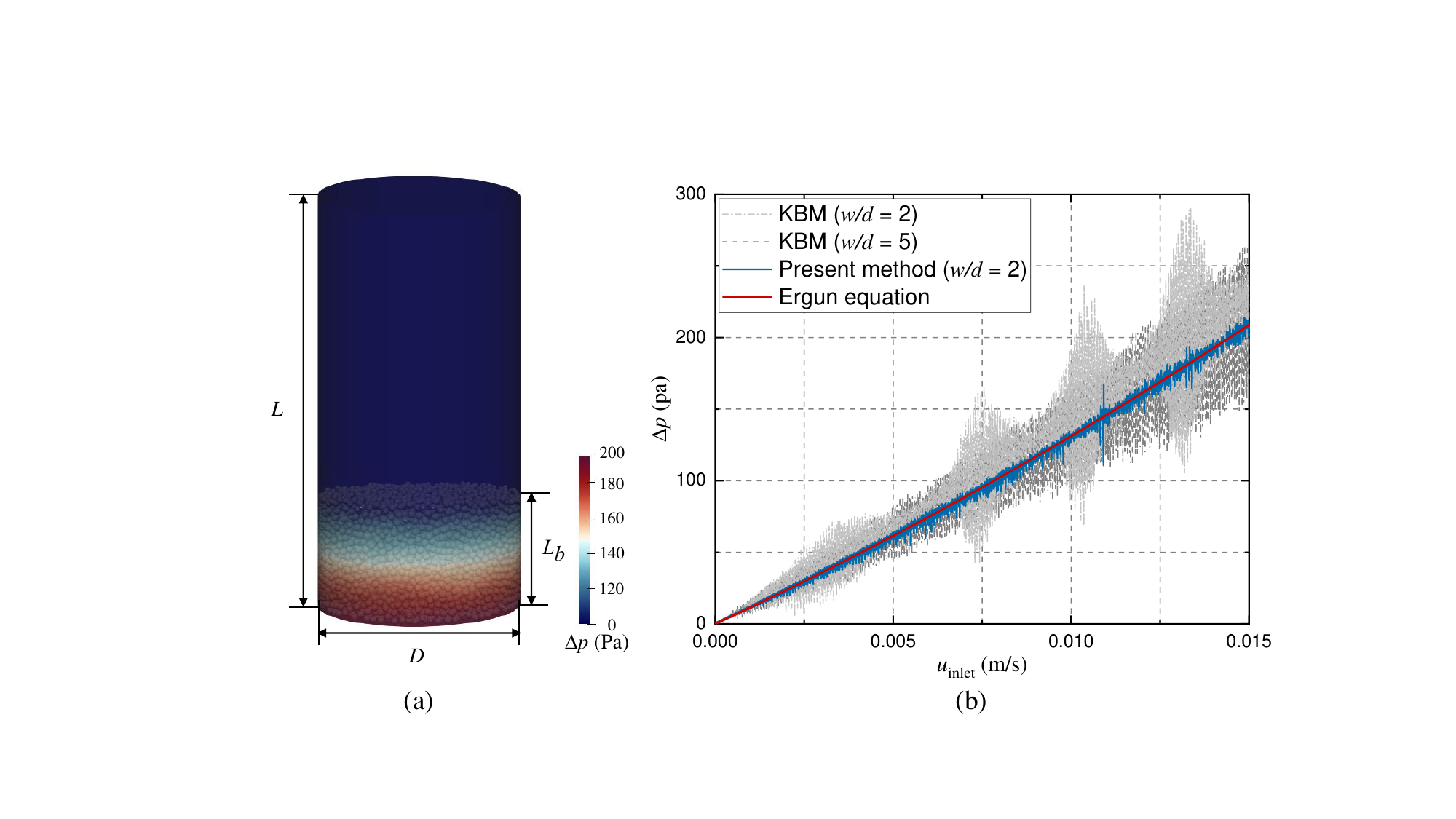}
     \caption{\label{fig:benchmarks_3d} Ergun test comparing the proposed method and KBM. (a) Computational domain. (b) Pressure drop as a function of the fluid inlet velocity.}
\end{figure}

The final test is a three-dimensional pressure drop test for a fluid flow through a dense granular bed in a cylinder column. As shown in Fig.~\ref{fig:benchmarks_3d}a, particles of diameter $d=0.001$~\si{m} and density $\rho_p=2000$~\si{kg/m^3} are packed at the bottom of a cylindrical column with diameter $D=28d $ and length $L = 56 d$. The granular bed is immobile with a thickness of $L_b=15d$ and volume fraction of $0.56$. To inject the fluid flow, a velocity inlet boundary condition with a linearly increasing value over time is applied at the bottom, while a fixed value pressure boundary condition is set at the top. Slip boundary conditions are imposed on the sides of the cylinder to eliminate sidewall effects on the pressure drop. The fluid injected from the bottom has a density $\rho_f=1000$~\si{kg/m^3} and viscosity $\mu_f=0.0015$~\si{Pa\cdot s}. We use a grid-to-particle size ratio, $\Delta x/d=1$, to test the proposed method and standard KBM with varying kernel width $w/d$ at a relatively fine grid resolution.

Figure~\ref{fig:benchmarks_3d}b shows the pressure drop $\Delta p$ as a function of the fluid inlet velocity $u_\textrm{inlet}$. The results are compared with the well-established Ergun equation (consistent with the drag model we use),
\begin{equation}\label{equ ergun test}
\Delta p=150\mu _fL_b\frac{\left( 1-\varepsilon _f \right) ^2u_\textrm{inlet}}{\varepsilon _{f}^{3}d^2}+1.75L_b\frac{1-\varepsilon _f}{\varepsilon _{f}^{3}}\frac{\rho _fu_\textrm{inlet}^2}{d}
\end{equation}
It is evident that, although both methods capture the overall increasing trend of $\Delta p$, the proposed method (with $w/d=2$) is accurate and has only small fluctuations comparing to the standard KBM (with $w/d=2$ and $5$). The significant fluctuations in the pressure field using the conventional method can be attributed to errors arising from the weight allocation, as tested above.

To summarize the three tests, the proposed method uses a point cloud-based mapping strategy to largely improve the accuracy, (temporal) stability and grid independence in the weight allocation processes from particles to grids, which is manifested in its good performance in the Ergun's test. On the contrary, the conventional kernel-based method that directly uses the grid-to-particle center distances to evaluate the weight function shows relatively poor accuracy and numerical stability, leading to significant pressure oscillations in Ergun's test using a fine grid configuration. Varying the kernel width and grid size can sometimes improve the results, but only the proposed method shows robustness when the grid resolution is adjusted.

\section{Validation by physical or numerical experiments}
In this section, the proposed method is validated by three physical or numerical experiments across dilute to dense particle-fluid systems, including sedimentation of a single particle, binary fluidized bed, and immersed granular column collapse.

\subsection{Sedimentation of a single sphere in a viscosity fluid}\label{validate:sedimentation}

\begin{figure}[htbp]
        \centering
        \includegraphics[width=\textwidth,height=6cm, keepaspectratio=true]{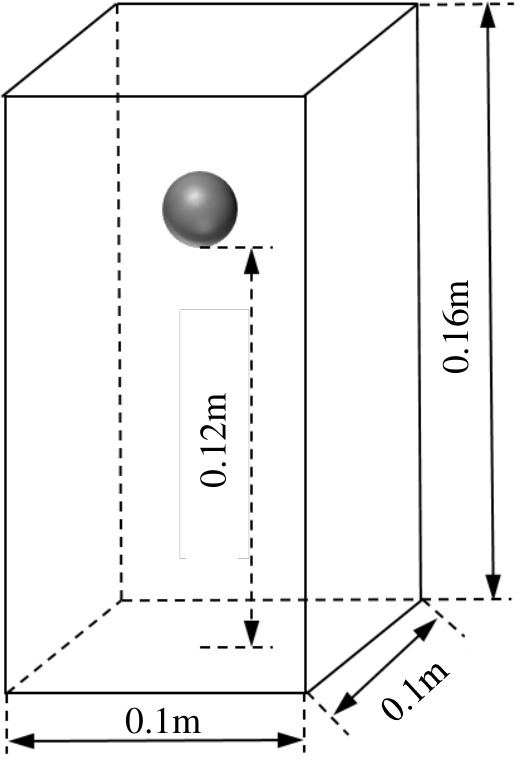}
     \caption{\label{fig:validation_single} Computational domain of the single sphere sedimentation test.}
\end{figure}

This test follows the experimental configuration of Cate et al.~\cite{ten2002particle}. As shown in Fig.~\ref{fig:validation_single}, the computational domain has a square base of $0.1$~\si{m} $\times$ $0.1$~\si{m} and a height of $0.16$~\si{m}. A nylon sphere with diameter $0.015$~\si{m} and density $1120$~\si{kg/m^3} is put $0.12$~\si{m} above the bottom and released with no initial velocity. The container is filled with silicon oil with a density of $960$~\si{kg/m^3} and dynamic viscosity of $0.058$~\si{Pa \cdot s}. The terminal settling velocity is predicted to be $0.128$~\si{m/s} in an infinite space. The virtual mass and time-related effects significantly influence the particle acceleration during the falling process, as the particle and fluid densities are similar~\cite{xie2021cfd}. Hence, a simple theoretical model proposed by Guo~\cite{guo2011motion} is adopted in this study to consider these effects by modifying the mass of the particle to $\frac{1}{6}\pi d_p^3\left( \rho _p+C_A\rho _f \right)$, where the coefficient $C_A =2$~\cite{wang2019semi,xie2021cfd}.

\begin{figure}[htbp]
        \centering
        \includegraphics[width=\textwidth,height=8cm, keepaspectratio=true]{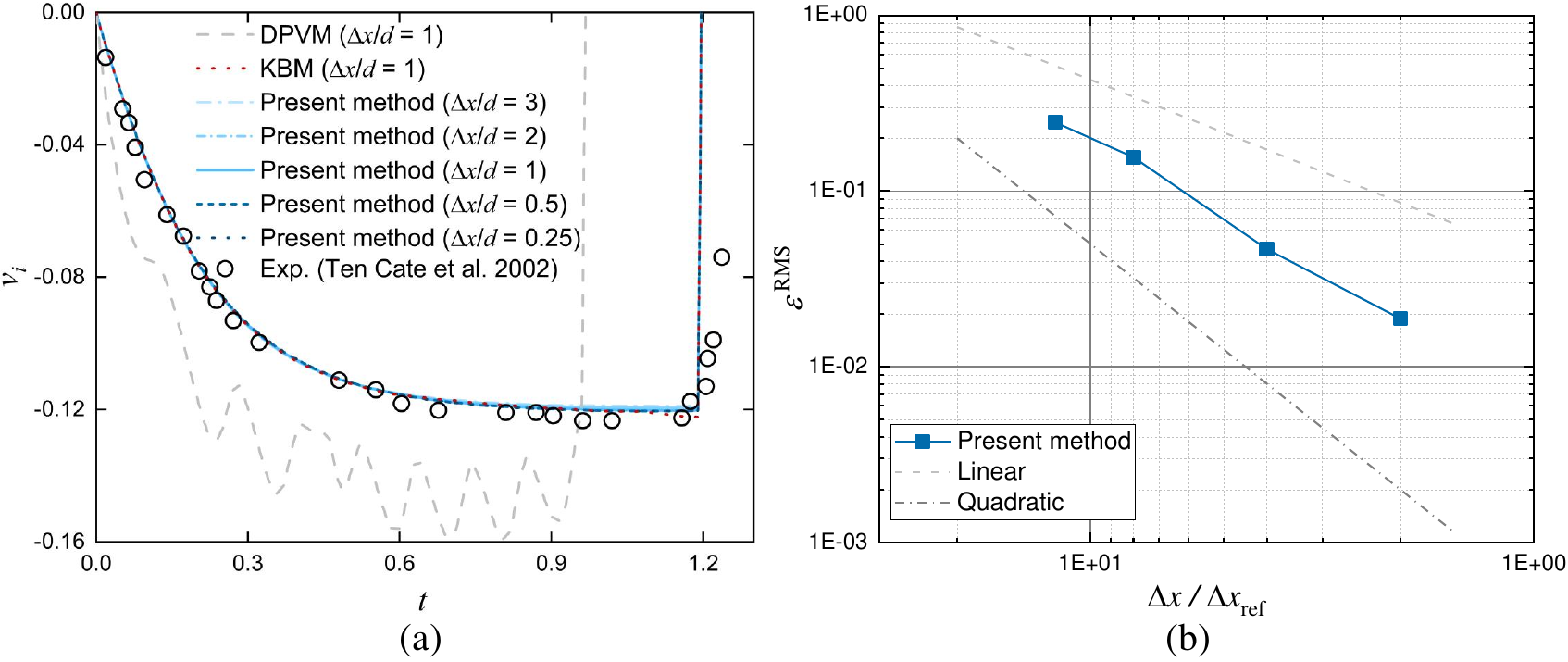}
        \caption{\label{fig:validation_single_results} Results of the single sphere sedimentation test. (a) Settling velocity obtained by the experiment, DPVM, and KBM with $\Delta x/d$ = 1, and the proposed method with $\Delta x/d$ = 3, 2, 1, 0.5, and 0.25. (b) Error convergence analysis for the proposed method.}
\end{figure}

The proposed method is validated in different grid configurations with the grid-to-particle size ratio $\Delta x/d = 3, 2, 1, 0.5$, and $0.25$. Figure~\ref{fig:validation_single_results}a plots the settling velocity, $v_i$, obtained by the proposed method, the conventional Divided Particle Volume Method (DPVM), and Kernel Based Method (KBM). The proposed method exhibits a good convergency under a twelve-fold change of grid size, whereas inaccurate and oscillating results are obtained from the DPVM when $\Delta x/d$ = 1. The velocity oscillation for DPVM is caused by the high gradient of the volume fraction and background velocity field due to the direct particle-to-local-grid mapping in a fine grid configuration. The results of the KBM with $w/d=2$ also align well with the experimental data, attributed to the distance-based weighting for particle-fluid interactions. Although its resolution is inherently grid-center-based, as previously mentioned, it proves to be sufficiently accurate for this particular problem.

In Fig.~\ref{fig:validation_single_results}b, we further perform a convergence analysis on the proposed method considering different $\Delta x/d$. The root-mean-square (RMS) error is used,
\begin{equation}\label{equ Errorconvergence}
\varepsilon ^{\text{RMS}}=\sqrt{\frac{1}{N_v}\sum_{i=1}^{N_v}{\left( \frac{v_i-v_{i,\textrm{ref}}}{v_{i,\textrm{ref}}} \right) ^2}}
\end{equation}
where $N_v = N_t/1000$ represents the sampling frequency during the sedimentation, which is chosen to ensure sufficient sampling accuracy, 
$N_t$ is the total time-steps, and the $v_{i,\text{ref}}$ and $\Delta x_{\text{ref}}$  are the reference velocity and grid resolution corresponding to the configuration of \(\Delta x/d = 0.25\). The dashed and dash-dotted lines represent the linear and quadratic convergence, respectively. It is demonstrated that the convergence of the settling velocity is stable with the refinement of the grids and the convergence rate of the proposed method exceeds the first order.

\subsection{Bi-disperse fluidized bed}

We use a bi-disperse fluidized bed test to validate the proposed method in relatively dilute but highly agitated particle-fluid systems where both coarse and fine particles are present. The results are compared with experiments~\cite{khan2016pressure} and simulations conducted using DPVM and KBM. The grid dependency of these methods is analyzed.

\begin{figure}[htbp]
        \centering
        \includegraphics[width=\textwidth,height=8cm, keepaspectratio=true]{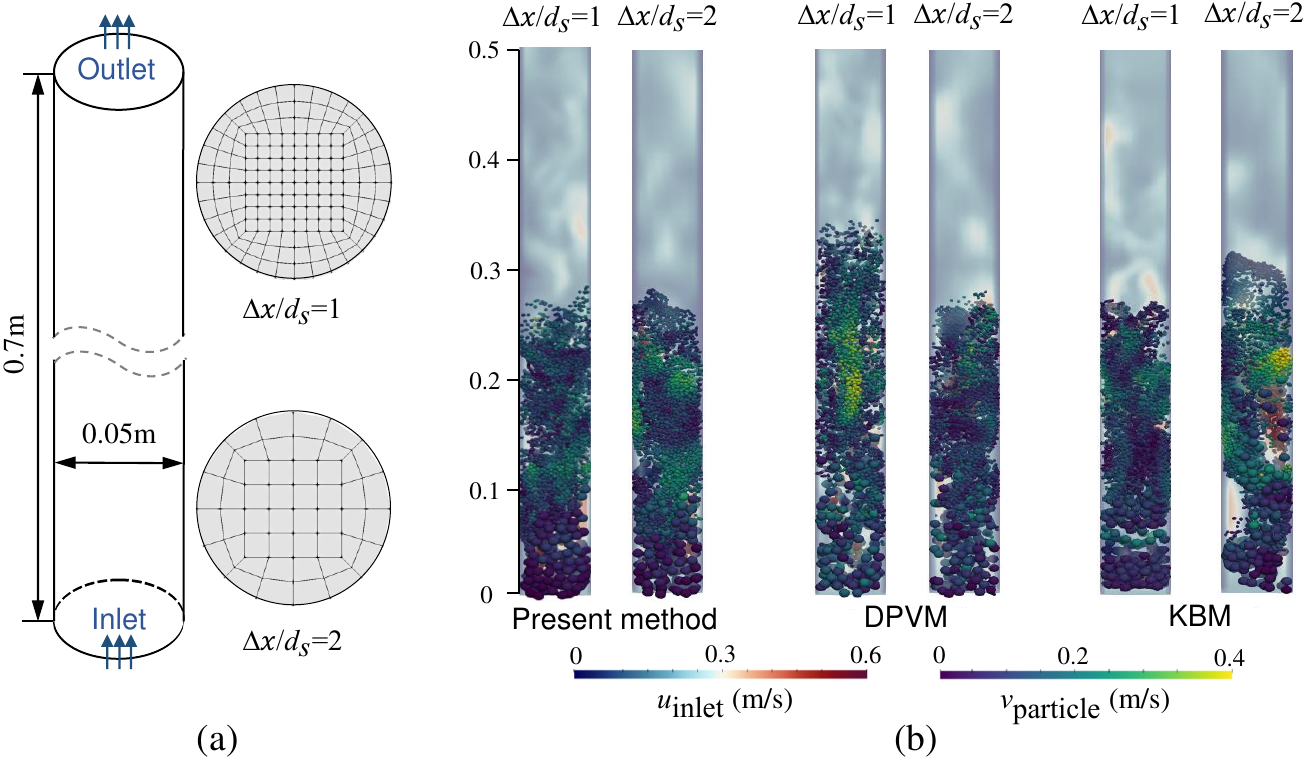}
     \caption{\label{fig:validation_bidisperse} Bi-disperse fluidized bed test. (a) Computational domain and grid configurations. (b) Comparison of the proposed method, DPVM, and KBM with different grid configurations under the fluid superficial velocity of $0.170$ ~\si{m/s}.}
\end{figure}

According to the experimental setup of Khan et al.~\cite{khan2016pressure}, the computational domain is a cylindrical vessel with a diameter of $0.05$~\si{m} and height of $0.7$~\si{m}. The diameters of large and small particles are $0.008$~\si{m} and $0.003$~\si{m}, respectively, leading to a size ratio of $2.67$. The large and small particle groups are each assigned a total mass of $0.12$~\si{kg}, and the particle density is set to be $2230$~\si{kg/m^3}. Due to the size difference, the number of particles in each group differs accordingly. The fluid density and dynamic viscosity are $1000$~\si{kg/m^3} and $0.001$~\si{Pa\cdot s}, respectively. The grid configuration is set to be fine ($\Delta x/d_s=1$ and $\Delta x/d_l= 0.37$) and coarse ($\Delta x/d_s=2$ and $\Delta x/d_l=0.75$), respectively, where $d_s$ and $d_l$ are the small and large particle diameters. The computational domain and the grid configurations are displayed in Fig.~\ref{fig:validation_bidisperse}a. 

We use five values of the inlet fluid superficial velocity $u_\textrm{inlet}$, namely, $0.141$, $0.149$, $0.156$, $0.163$, and $0.170$~\si{m/s}, consistent with the experiments. Figure~\ref{fig:validation_bidisperse}b shows typical snapshots for $u_\textrm{inlet}=0.170$~\si{m/s}. It is evident that the simulation results of the proposed method are very similar for both fine and coarse grids, whereas the results of DPVM and KBM show a strong dependence on the grid configuration.

\begin{figure}[htbp]
        \centering
        \includegraphics[width=\textwidth,height=8cm, keepaspectratio=true]{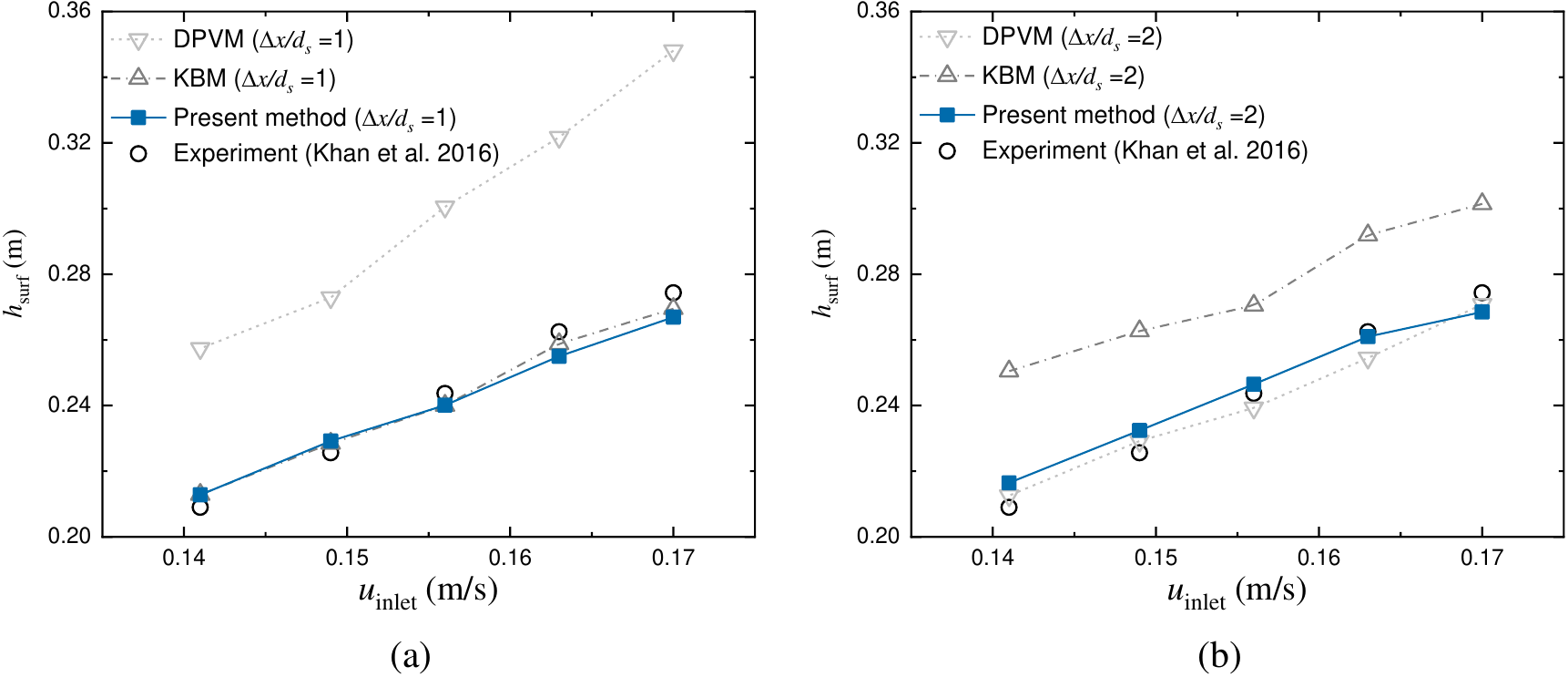}
    \caption{\label{fig:validation_bidisperse_h} Steady-state expansion heights $h_\textrm{surf}$ of the fluidized bed under different fluid velocities $u_\textrm{inlet}$ with (a) $\Delta x/d_s$ = 1 and (b) $\Delta x/d_s$ = 2.}
\end{figure}

We further extract the steady state time-averaged expansion height $h_\textrm{surf}$ for various $u_\textrm{inlet}$ to compare the performance of different methods. As shown in Fig.~\ref{fig:validation_bidisperse_h}, for fine grids, both KBM and the proposed method exhibit a good agreement with the experimental data, but the results of DPVM is significantly higher. This can be attributed to the high gradients of field quantities (e.g., void fraction, grid-based fluid and particle velocities) typical of DPVM for fine grids~\cite{wang2019semi,chen2022semi,eshraghi2023coarse}. For coarse grids, DPVM shows an improved result but KBM overestimates $h_\textrm{surf}$ for all values of $u_\textrm{inlet}$.  The errors in KBM for coarser grids can be attributed to the increased errors of weight allocation, as discussed in Section~\ref{sec:benchmarks_static}. In both fine and coarse grid configurations, our proposed method using a two-step mapping strategy produces accurate results for the bi-disperse system, because the weight allocation and the subsequent computations are independent of the grid-to-particle size ratio. This highlights the robustness of the proposed method when varied particle sizes and grid configurations are encountered. 

\newpage\subsection{Immersed granular column collapse with varying initial packing densities}\label{sec:immersed granular collapse}

The final validation test is regarding the collapse of an immersed granular column. It has been widely reported that, unlike dry granular collapse, the initial column packing density, $\phi_i$, plays a significant role in this scenario. The central idea is that granular materials tend to reach a specific, critical volume fraction ($\phi_{cr}$) during shear deformation. In a dense granular packing, $\phi_i>\phi_{cr}$, shear dilation occurs and leads to negative excess pore fluid pressure, which enhances the frictional resistance of the granular packing according to the effective stress principle in soil mechanics. Such pore pressure feedback largely decelerates the initiation and propagation of the immersed granular flow. On the contrary, a loose packing  $\phi_i<\phi_{cr}$ leads to shear contraction and hence acceleration of the immersed granular collapse. It is noteworthy that naturally formed granular columns usually have a packing density between $0.54$ and $0.6$, and the critical packing density is around $0.58$. Moreover, even subtle volume changes of the granular phase can lead to significant excess pore fluid pressure and different collapse dynamics~\cite{iverson2000acute,pailha2009two,rondon2011granular}. As such, it is crucial for the CFD-DEM algorithm to accurately capture the changes in the granular volume fraction and their contribution to the fluid-particle interaction force in this test.

\begin{figure}[htbp]
        \centering
        \includegraphics[width=\textwidth,height=4.5cm, keepaspectratio=true]{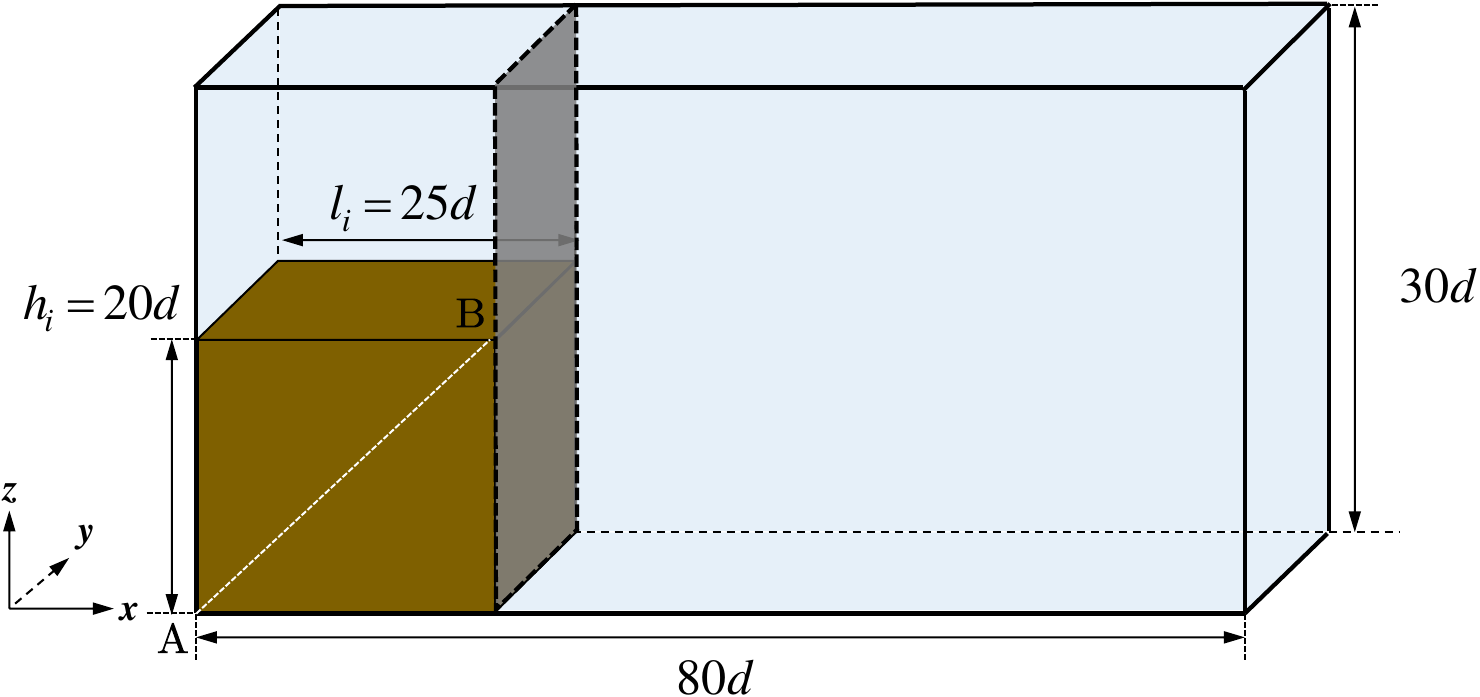}
     \caption{\label{fig: Geometry of granular collapse} Computational setup of the immersed granular collapse test.}
\end{figure}

To allow a detailed assessment of the proposed method in terms of the internal pore pressure fields, we use fully resolved lattice boltzmann method-discrete element method (LBM-DEM) simulations conducted by Yang et al.~\cite{yang2020pore} for validation. The numerical setup is shown in Fig.~\ref{fig: Geometry of granular collapse}. The particle diameter and density are $d = 1~\text{mm}$ and $\rho_p = 2500~\text{kg/m}^3$, respectively. The initial length and height of the granular column are $l_i = 25d$ and $h_i = 20d$, respectively, fully immersed in a viscous fluid with a viscosity of $\mu_f = 0.01~\text{Pa·s}$ and a density of $\rho_f = 1000~\text{kg/m}^3$. The computational domain is $ 80d \times 10d \times 30d$ in dimensions, with no-slip boundary conditions imposed at $z = 0$, $x = 0$, and $x = 80d$. A slip boundary condition is applied at the top surface, while periodic boundary conditions are defined in the $y$-direction for both the particle and fluid phases. We follow the generation procedure of the granular column in Yang et al.~\cite{yang2020pore}. However, the maximum value reported in Yang et al.~($\phi_i = 0.6277$) could not be exactly reproduced in our simulations. Nevertheless, this slight discrepancy does not affect the overall trend and the conclusions.

\begin{figure}[htbp]
        \centering
\includegraphics[width=\textwidth,
        height=10cm,
        keepaspectratio=true]{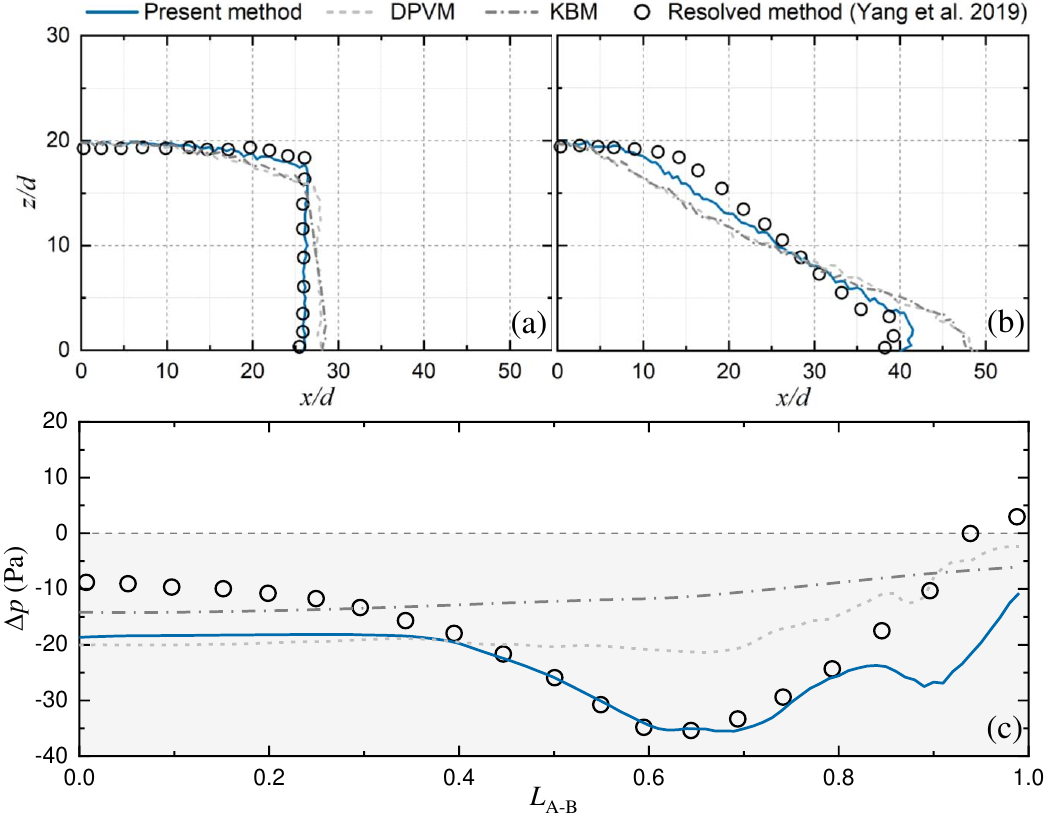}
     \caption{\label{fig: granular surface and pore pressure} Results of a very dense granular column with $\phi_i=0.6233$. (a) and (b) compare the granular free surface at $t=0.1$ s and $t=0.4$ s, respectively, for different methods. (c) Temporal and spatial averaged fluid pore pressure along the diagonal plane A-B (indicated by a normalized location from A to B, $L_{A-B}$) of the granular column during the initiation stage ($t=0 \sim 0.06$ s).}
\end{figure}

We first use a dense case, $\phi_i=0.6233$, to demonstrate the comparison of the free surfaces  at two different time instances. As shown in Figs.~\ref{fig: granular surface and pore pressure}a and b, the results obtained from the resolved method show that the granular column has nearly no displacement at $t=0.1$\si{s} due to the pore pressure feedback of shear dilation. The proposed CFD-DEM accurately captures this behavior, while the conventional DPVM and KBM predict a much faster collapse at this time instance likely because they fail to capture the pore pressure induced by subtle displacements in the granular packing. At a later time instance, $t=0.4$\si{s}, only our proposed method accurately captures the free surface of the granular flow. 
Furthermore,  Fig.~\ref{fig: granular surface and pore pressure}c presents the fluid pore pressure during the beginning of the collapse along the diagonal plane of the granular column $L_{\text{A-B}}$ (i.e., section A-B in Fig.~\ref{fig: Geometry of granular collapse}), consistent with the sampling location used in the reference study~\cite{yang2020pore}. Note that each point in Fig.~\ref{fig: granular surface and pore pressure}c denotes a temporal average for $t=0\sim0.06$~\si{s} and a spatial average along the $y$-direction. It is clear that the fluid pore pressures predicted by KBM and DPVM are indeed lower than the resolved results, indicating that the dilation-induced pore pressure generation is not captured, whereas the proposed method predicts the negative pore pressure reasonably well, especially near the top-right corner of the granular column where the particle motion is initiated. 

\begin{figure}[htbp]
        \centering
        \includegraphics[width=\textwidth,height=7cm, keepaspectratio=true]{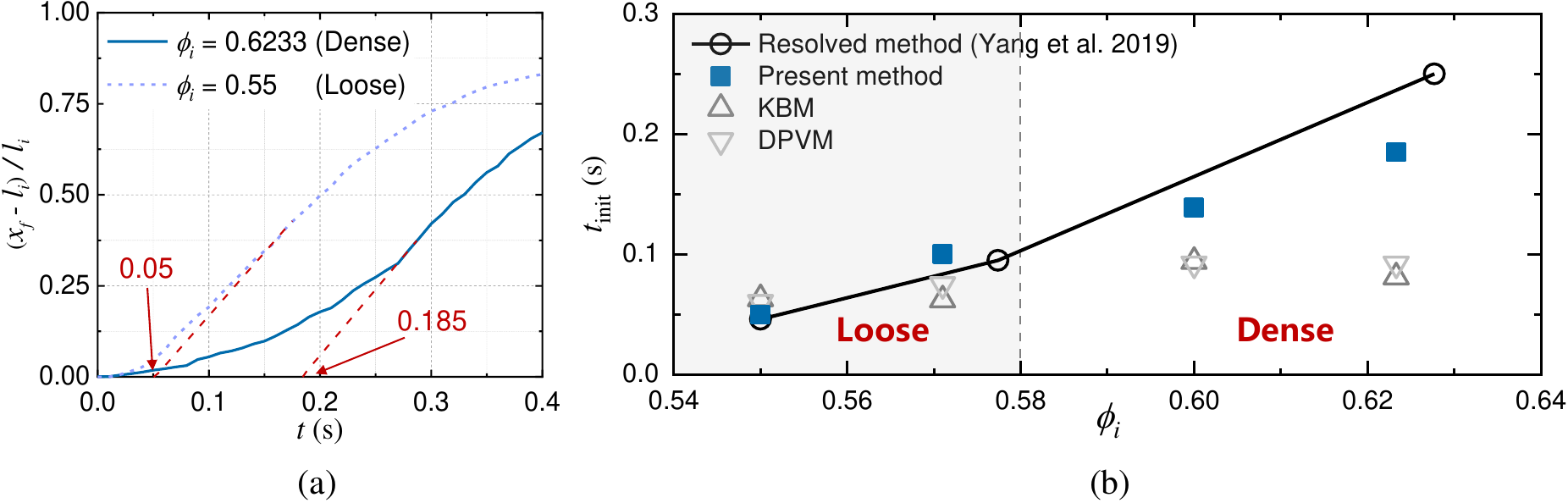}
     \caption{\label{fig valid3initime} Results of varying initial packing densities $\phi_i$. (a) Temporal evolution of the frontal position $x_f$ for a dense and a loose packing configurations, illustrating the definition of the initiation time $t_\textrm{init}$. (b) $t_\textrm{init}$ as a function of $\phi_i$ for columns having loose to very dense initial states, comparing different methods.}
\end{figure}

Next, we focus on the initiation time for a wide range of initial packing densities, $\phi_i=0.55, 0.571, 0.60, 0.6233$. The initiation time $t_\textrm{init}$ is defined as the intercept of the linear fit of the maximum slope in the runout profile with the time axis, consistent with the reference~\cite{yang2020pore}. 
Figure~\ref{fig valid3initime}a provides a detailed illustration of how the initiation time is identified, with the vertical axis showing the normalized runout distance, $(x_f-l_i)/l_i$, where $x_f$ is the front position in the $x$-direction, and the horizontal axis indicating the time of initiation. It can be observed that the proposed method captures the distinct dynamic behaviors of collapse in dense and loose packings.
Then, we compare the proposed and conventional methods with the initiation times obtained from the resolved LBM-DEM method~\cite{yang2020pore} in Fig.~\ref{fig valid3initime}b.
The results obtained by the proposed method closely follow the trend of the resolved method. As the particle volume fraction increases, a noticeable delay in collapse initiation is captured. This delay is attributed to sufficiently negative pore pressure, which has been confirmed as a result of the dense initial packing~\cite{pailha2008initiation,pailha2009two,rondon2011granular}.
Capturing such a phenomenon requires detecting the sub-grid displacements of particles and their impact on phase momentum exchange at the beginning of the collapse dynamics, which clearly demonstrates the advantages of the proposed two-step mapping strategy. 

In summary, the three validation cases demonstrate that the volume-averaged CFD-DEM framework can be enhanced by the proposed method for a better prediction of dilute to dense particle-fluid systems, with evident advantages in coupling accuracy and grid-independence under both fine- and coarse-grid configurations. Notably, only the proposed approach accurately captures the pore pressure feedback governing the onset of very dense granular column collapse.

\section{Efficiency analysis}\label{sec: efficiency}
\begin{figure}[htbp]
        \centering
        \includegraphics[width=\textwidth,height=7cm, keepaspectratio=true]{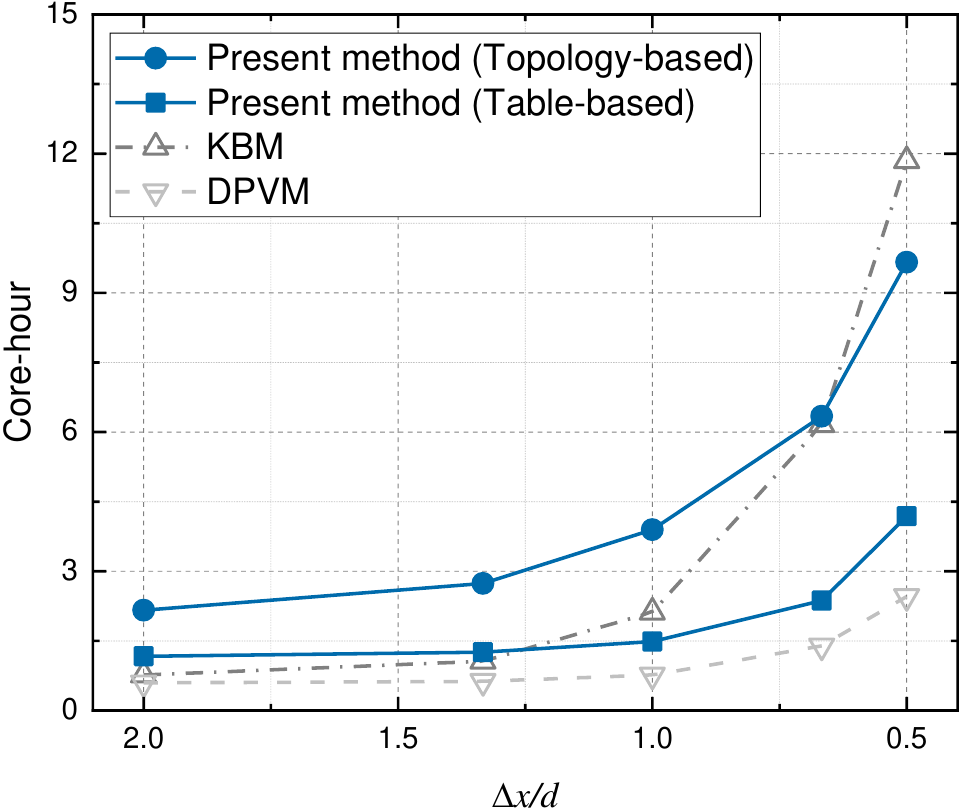}
     \caption{\label{fig:efficiency} Efficiency test for the proposed method, with both topology-based and table-based point allocation strategies, and conventional methods.}
\end{figure}

In Fig.~\ref{fig:efficiency}, we perform an efficiency analysis for the proposed method, compared to conventional methods, based on the immersed granular collapse test (case $\phi_i=0.6233$ in Section~\ref{sec:immersed granular collapse}). The configurations with varying grid sizes are adopted to test the performance among DPVM, KBM, and the proposed method. The size ratio of grid to particle $\Delta x/d$ is 2, 1.33, 1, 0.67, and 0.5, corresponding to total grid numbers of 3000, 11040, 24000, 81000, 192000, respectively. DPVM consistently exhibits the highest computational efficiency across all configurations, as it does not require any particle- or grid-based searching operations. 
At coarse grid resolutions (i.e., with fewer grids), the computational efficiency of KBM is comparable to DPVM and exceeds the proposed method. However, as the grid is refined, its efficiency decreases rapidly. For $\Delta x/d\ge 1.5$, the efficiency of KBM falls below that of the proposed method. 
The KBM involves searching between particles and grid centers, which significantly increases the computational cost in fine-grid configurations. 
The proposed method eliminates such searching by introducing a topology identification between particles and grid surfaces, which allows for improved efficiency under fine grid configurations.
However, this approach requires additional processing to construct the point cloud representation, resulting in slightly higher overhead on coarse grids compared to KBM.

To further improve the efficiency of our proposed method, a look-up table-based strategy is developed to replace the original topology identification in structured grid configurations. The main idea of the table-based method lies in pre-storing the correspondence between spatial coordinates and grid IDs in a look-up table, avoiding costly distance- or topology-based calculation between point clouds and grids. By doing so, the table-based method achieves advanced computational efficiency across all grid configurations (Fig.~\ref{fig:efficiency}). Moreover, this strategy is inherently parallelizable and can be easily integrated into existing CFD-DEM frameworks.

\section{Conclusions}

This study presents a novel mapping strategy that integrates the concept of point-based coarse graining into the volume-averaged CFD-DEM framework, enabling accurate simulation of dilute to dense particle-fluid systems, with particular effectiveness in densely packed granular configurations. The proposed mapping strategy consists of two steps. In the first step, discrete particles are converted into grid-independent coarse-grained fields via the Fibonacci point cloud. Then, the coarse-grained fields are coupled with the fluid fields using a point-based projection, accounting for accurate two-way interaction force coupling. 

A series of tests are conducted to demonstrate the enhanced performance and robustness of the proposed algorithm.
The proposed method is first tested against three idealized numerical tests, demonstrating superior accuracy and robustness compared to the conventional grid-based mapping method (KBM). Further, it is validated against three published experimental and fully-resolved numerical data, ranging from single particle, loose fluidized, to dense packing configurations. These results suggest that the proposed strategy enables an accurate detection of subtle particle displacement and their influence on pore pressure feedback, which has been widely recognized as a key mechanism governing the dynamics of dense particle-fluid systems. 
In addition, this strategy is able to provide grid-independent results over a wider range of particle-to-grid size ratios, especially on a relatively coarse grid configuration, compared to the standard KBM. 

The proposed method provides a unified and robust framework for simulating particle–fluid interactions across a wide range of particle-to-grid size ratios and solid volume fractions.
The proposed method can be used to investigate the failure, rheology, and segregation of immersed granular systems, which typically require long-term and large-scale simulations.
In addition, future work will focus on extending the coarse graining strategy to non-spherical particle–fluid systems, with the aim of broadening its applicability to complex industrial processes and geophysical flows. 

\section*{Acknowledgements}
This research is supported by the National Natural Science Foundation of China (Grant no. 12472412). H.B. Shi acknowledges the financial support provided by the Science and Technology Development Fund, Macau S.A.R. (File no. 001/2024/SKL).

\section*{CRediT authorship contribution statement}
Y.X. Liu: Investigation, Methodology, Software, Data curation, Writing – original draft, Formal analysis.  
L. Jing: Conceptualization, Investigation, Methodology, Data curation, Writing – review \& editing, Formal analysis, Supervision, Funding acquisition.
X.D. Fu: Conceptualization, Supervision, Funding acquisition.
H.B. Shi: Conceptualization, Methodology, Writing – review \& editing, Funding acquisition.  

\section*{Declaration of competing interest}
No potential conflict of interest was reported by the authors.

\appendix
\section{Point-based treatments for the non-physical and physical boundaries} \label{app:boundary}
We introduce the treatments of the non-physical (e.g., processor and periodic) and physical boundaries (e.g., walls) for the proposed CFD-DEM method via point-based operation. Different boundary conditions can be managed by simply adjusting the positions and weights of the point-based field instead of extra searching and renormalization weighting.
\par
\begin{figure}[htbp]
        \centering
        \includegraphics[width=\textwidth,height=5cm, keepaspectratio=true]{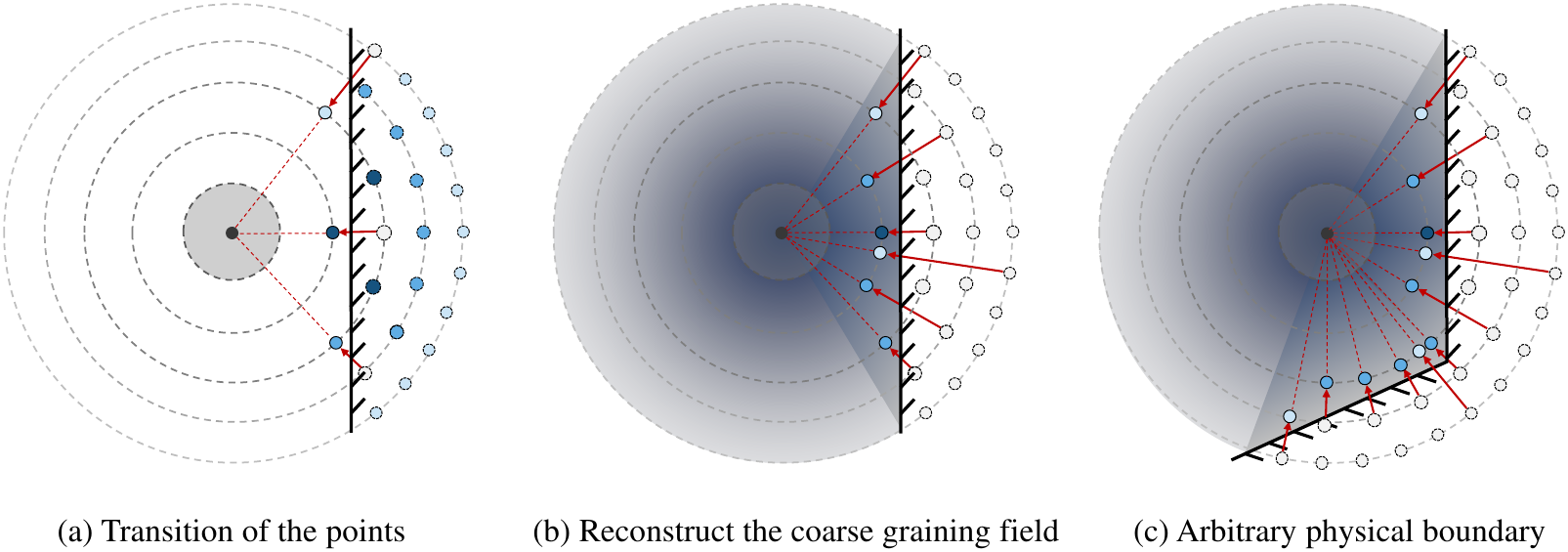}
     \caption{\label{fig:physicalboundary} No-flux point approach for physical boundaries.}
\end{figure}

 For non-physical boundaries, the implementation is straightforward, as it involves exchanging point cloud information without the need for grid-based searching across processors. For processor boundaries, local point positions and weights are first collected within each processor and then gathered at the main processor. The main processor subsequently scatters the corrected positions and weights back to their respective processors through a reduction operation. For periodic boundaries, point coordinates are directly transformed according to the periodic conditions, eliminating the need for grid-based searching. When particles are near the physical boundaries, expansion domains may extend beyond the computational domain, requiring special treatments to preserve conservation laws such as dynamically adjusting the renormalization coefficient~\cite{xiao2011algorithms}, image kernel method~\cite{ries2014coarse,eshraghi2023coarse,zhu2002averaging}, and zero-gradient boundary method for diffusion-based method~\cite{sun2015diffusion}. Inspired by the work of Sun et al.~\cite{sun2015diffusion}, a no-flux point approach is proposed in this work for the physical boundaries. In this method, when points extend beyond the physical boundary, they are retracted to the previous layer of the point cloud, gradually moving toward the particle center until they are positioned within the nearest interior layer of the domain (see Fig.~\ref{fig:physicalboundary}a). In this manner, mass conservation is inherently ensured by maintaining the weights of the points unchanged during the adjustment process. The coarse graining field is then reconstructed by the updated position of points (see Fig.~\ref{fig:physicalboundary}b). A notable advantage of this strategy is its natural suitability for arbitrary physical boundaries (see Fig.~\ref{fig:physicalboundary}c), as the adjustments are directly based on the relative positions between sub-points and their corresponding particle centers, inherently capturing the influence of complex boundary geometries without requiring additional modifications.

\par
\begin{figure}[htbp]
        \centering
        \includegraphics[width=\textwidth,height=10cm, keepaspectratio=true]{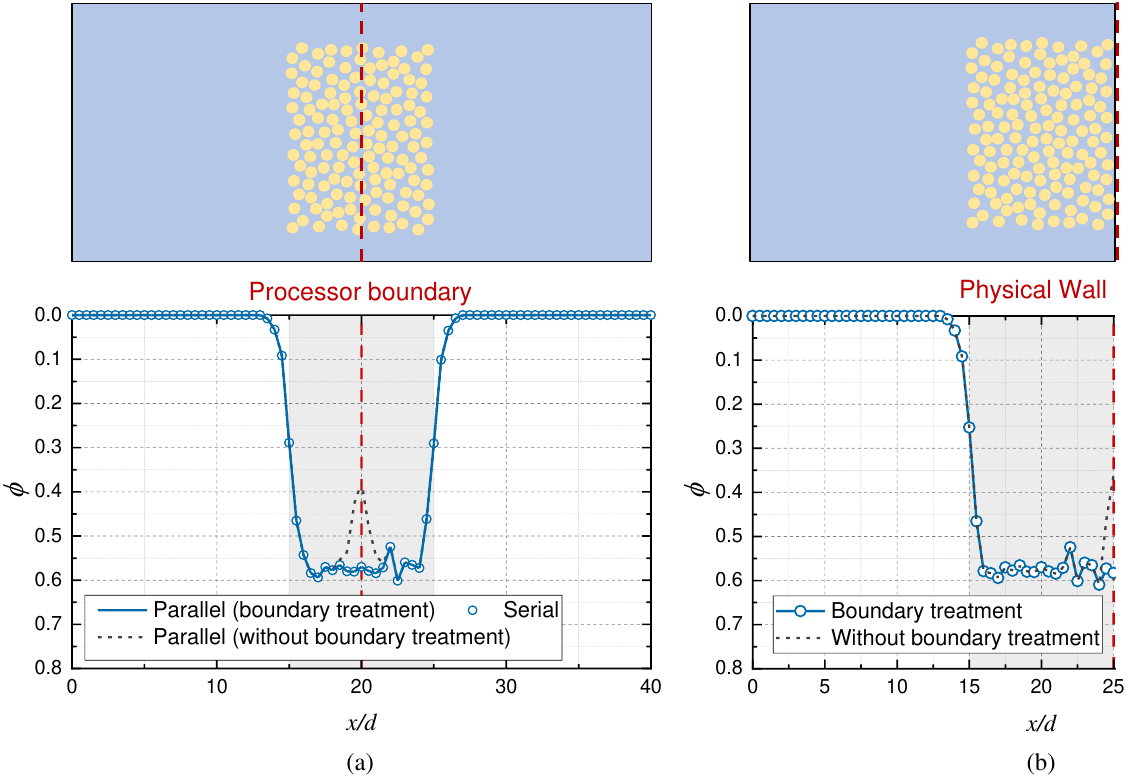}
         \caption{\label{fig:volumefractiontest_boundarytreatment} Volume fraction distribution of a natural densely packed granular column near (a) non-physical and (b) physical boundaries.}
\end{figure}

 To examine the proposed treatment, we evaluated the volume fraction distribution of a densely packed granular column near both non-physical and physical boundaries, where the particle volume fraction reaches approximately 58\%~\cite{rondon2011granular,pailha2008initiation}. Figure~\ref{fig:volumefractiontest_boundarytreatment} shows the volume fraction profiles along the sampling line through the column's centroid, considering (a) processor boundaries in parallel computing and (b) physical wall boundaries. The red dashed lines indicate the processor or physical boundaries.

For parallel computing, the proposed method achieves consistency with direct serial computation, effectively preserving particle information across processor boundaries. In contrast, neglecting specific boundary treatments causes significant information loss at processor interfaces. Similarly, near physical walls, our method accurately captures the volume fraction distribution, showing a good agreement with theoretical and experimental results. In the absence of appropriate boundary handling, particle information near walls is lost, leading to violations of mass conservation.

\par\section{Stability criterion for dense particle-fluid systems and an extended semi-implicit momentum exchange scheme using coarse-grained velocity} \label{app:stability}

In the explicit momentum exchange term of the CFD-DEM governing equations, a basic stability criterion is that the coupling time-step should be limited by the two-way interaction force governed by Newton's laws~\cite{blais2016development}. Since the drag force is regarded as the primary component of the interaction force, considering the relative inertia of particle and fluid phase, the time-step criterion can be derived as:
\begin{equation}
\Delta t_p<\frac{\boldsymbol{u}-\boldsymbol{v}}{\boldsymbol{a}_p}=\frac{\boldsymbol{u}-\boldsymbol{v}}{\boldsymbol{f}_{drag}/\left( V_p\rho _p \right)}
\end{equation}
\begin{equation}
\Delta t_f<\frac{\boldsymbol{u}-\boldsymbol{v}}{\boldsymbol{a}_f}=\frac{\boldsymbol{u}-\boldsymbol{v}}{\boldsymbol{f}_{drag}/\left( V_p\rho _f \varepsilon_f \left( 1-\varepsilon_f \right) \right)}
\end{equation}
where $\boldsymbol{a}_p$ and $\boldsymbol{a}_f$ are the particle and fluid accelerations, respectively, $\boldsymbol{u}$ is the fluid velocity, $\boldsymbol{v}$ is the particle velocity, and $\Delta t_p$ and $\Delta t_f$ are the time-step limitations on particle and fluid, respectively. Substituting Eq.~\eqref{equ gidaspow drag} into the above expression, yields:
\begin{equation}\label{Equ taoP}
\Delta t_p<\frac{\left( 1-\varepsilon_f \right) \rho _p}{\beta}
\end{equation}
\begin{equation}\label{Equ taoF}
\Delta t_f<\frac{\varepsilon_f \rho _f}{\beta}
\end{equation}
where $\beta$ denotes the drag coefficient in the Gidaspow drag model (see Eq.~\eqref{equ:gidaspow-drag-coe}). Considering a single sphere in the Stokes regime, the Eq.~\eqref{Equ taoP} reduces to the particle relaxation time $\Delta t_p=d_{p}^{2}\rho _p/\left( 18\mu _f \right)$.
\par In a dense particle–fluid system, by substituting the Ergun-type coefficient \(\beta\) into Eq.~\eqref{Equ taoF} and rearranging the resulting terms, the following stability criterion can be derived:
\begin{equation}\label{Equ taoPErgun}
\Delta t_f<\frac{1}{A\left( \frac{1-\varepsilon_f}{\varepsilon_f} \right) ^2+B\left( \frac{1-\varepsilon_f}{\varepsilon_f} \right)}
\end{equation}
where $A=150\mu _f/\left( \rho _fd^2 \right) $, $B=1.75\left| \boldsymbol{u}-\boldsymbol{v} \right|/d$. We have now established the stability criterion related to the volume fraction. Note that similar results can be obtained using alternative drag force models~\cite{blais2016development}. Equation~\eqref{Equ taoPErgun} highlights the importance of adopting a stable momentum exchange strategy in dense particle-fluid systems~\cite{blais2016development}.

\par Moreover, previous studies have shown that separating the momentum exchange term into implicit and explicit terms greatly improves numerical stability~\cite{kloss2012models,radl2015state} in CFD-DEM. To enable such a semi-implicit treatment within the volume-averaged N-S equations, the discrete particle velocity must be accurately projected onto a continuous representation. Building on the semi-implicit framework proposed by Goniva et al.~\cite{goniva2015open,radl2015state}, we incorporate our two-step coarse graining strategy on the particle velocity field to improve numerical stability in dense particle-fluid systems.
The explicit particle-to-fluid interaction term in the fluid governing equations (see Eq.~\eqref{equ force to fluid}) can be reformulated into an extended semi-implicit form using the proposed coarse-grained velocity field (see Eq.~\eqref{eq:gridbased particle-velocity}), which can be written as:
\begin{equation}
\left< \boldsymbol{F}_{p-f} \right> _j=\left< \frac{\left| \sum_{i=1}^{N_p}{\left( \boldsymbol{f}_{drag,i}\mathcal{W}_{i,j} \right)} \right|}{V_j\left| \boldsymbol{u}-\left< \boldsymbol{v} \right> _j \right|}\boldsymbol{u} \right> _{\text{impl}}-\left< \frac{\left| \sum_{i=1}^{N_p}{\left( \boldsymbol{f}_{drag,i}\mathcal{W}_{i,j} \right)} \right|}{V_j\left| \boldsymbol{u}-\left< \boldsymbol{v} \right> _j \right|}\left< \boldsymbol{v} \right> _j \right> _{\exp\text{l}}
\end{equation}
where the $\left< \boldsymbol{F}_{p-f} \right> _j$ is the semi-implicit particle-to-fluid interaction term, $N_p$ is the number of particles whose expansion domains interact with grid $j$, 
$\mathcal{W}_{i,j}$ is the normalized weights assigned to grid $j$ from particle $i$, $\boldsymbol{f}_{drag,i}$ is the drag force acting on particle $i$,  
$V_j$ is the volume of the grid $j$, and $\boldsymbol{u}$ and $\left<\boldsymbol{v}\right>_{j}$ are the fluid and coarse-grained particle velocity at the grid $j$, respectively. The first term on the right-hand side, representing the contribution of fluid velocity, is handled implicitly, while the second term, associated with particle velocity, is treated explicitly.

\bibliographystyle{elsarticle-num}
\bibliography{cite}

\begin{thebibliography}{10}
\expandafter\ifx\csname url\endcsname\relax
  \def\url#1{\texttt{#1}}\fi
\expandafter\ifx\csname urlprefix\endcsname\relax\def\urlprefix{URL }\fi
\expandafter\ifx\csname href\endcsname\relax
  \def\href#1#2{#2} \def\path#1{#1}\fi

\bibitem{talling2007onset}
P.~Talling, R.~Wynn, D.~Masson, M.~Frenz, B.~Cronin, R.~Schiebel, A.~Akhmetzhanov, S.~Dallmeier-Tiessen, S.~Benetti, P.~Weaver, et~al., Onset of submarine debris flow deposition far from original giant landslide, Nature 450 (2007) 541--544.

\bibitem{deal2023grain}
E.~Deal, J.~G. Venditti, S.~J. Benavides, R.~Bradley, Q.~Zhang, K.~Kamrin, J.~T. Perron, Grain shape effects in bed load sediment transport, Nature 613 (2023) 298--302.

\bibitem{peacock2023fluid}
T.~Peacock, R.~Ouillon, The fluid mechanics of deep-sea mining, Annu. Rev. Fluid Mech. 55 (2023) 403--430.

\bibitem{shi2021theoretical}
H.~Shi, P.~Dong, X.~Yu, Y.~Zhou, A theoretical formulation of dilatation/contraction for continuum modelling of granular flows, J. Fluid Mech. 916 (2021) A56.

\bibitem{topin2012collapse}
V.~Topin, Y.~Monerie, F.~Perales, F.~Radjai, Collapse dynamics and runout of dense granular materials in a fluid, Phys. Rev. Lett. 109 (2012) 188001.

\bibitem{ishikawa2022lubrication}
T.~Ishikawa, Lubrication theory and boundary element hybrid method for calculating hydrodynamic forces between particles in near contact, J. Comput. Phys. 452 (2022) 110913.

\bibitem{dance2003incorporation}
S.~Dance, M.~Maxey, Incorporation of lubrication effects into the force-coupling method for particulate two-phase flow, J. Comput. Phys. 189 (2003) 212--238.

\bibitem{boyer2011unifying}
F.~Boyer, {\'E}.~Guazzelli, O.~Pouliquen, Unifying suspension and granular rheology, Phys. Rev. Lett. 107 (2011) 188301.

\bibitem{blais2016development}
B.~Blais, M.~Lassaigne, C.~Goniva, L.~Fradette, F.~Bertrand, Development of an unresolved {CFD--DEM} model for the flow of viscous suspensions and its application to solid--liquid mixing, J. Comput. Phys. 318 (2016) 201--221.

\bibitem{rondon2011granular}
L.~Rondon, O.~Pouliquen, P.~Aussillous, Granular collapse in a fluid: role of the initial volume fraction, Phys. Fluids 23 (2011).

\bibitem{pailha2008initiation}
M.~Pailha, M.~Nicolas, O.~Pouliquen, Initiation of underwater granular avalanches: Influence of the initial volume fraction, Phys. Fluids 20 (2008).

\bibitem{jop2006constitutive}
P.~Jop, Y.~Forterre, O.~Pouliquen, A constitutive law for dense granular flows, Nature 441 (2006) 727--730.

\bibitem{guazzelli2018rheology}
{\'E}.~Guazzelli, O.~Pouliquen, Rheology of dense granular suspensions, J. Fluid Mech. 852 (2018) P1.

\bibitem{sun2023two}
Y.~Sun, J.~Li, Z.~Cao, A.~G.~L. Borthwick, A two-dimensional double layer-averaged model of hyperconcentrated turbidity currents with non-newtonian rheology, Int. J. Sediment Res. 38 (2023) 794--810.

\bibitem{malkus1990dynamics}
D.~S. Malkus, J.~A. Nohel, B.~J. Plohr, Dynamics of shear flow of a non-newtonian fluid, J. Comput. Phys. 87 (1990) 464--487.

\bibitem{hu2001direct}
H.~H. Hu, N.~A. Patankar, M.~Zhu, Direct numerical simulations of fluid--solid systems using the arbitrary {Lagrangian--Eulerian} technique, J. Comput. Phys. 169 (2001) 427--462.

\bibitem{peskin1977numerical}
C.~S. Peskin, Numerical analysis of blood flow in the heart, J. Comput. Phys. 25 (1977) 220--252.

\bibitem{feng2005proteus}
Z.-G. Feng, E.~E. Michaelides, Proteus: a direct forcing method in the simulations of particulate flows, J. Comput. Phys. 202 (2005) 20--51.

\bibitem{yu2006fictitious}
Z.~Yu, X.~Shao, A.~Wachs, A fictitious domain method for particulate flows with heat transfer, J. Comput. Phys. 217 (2006) 424--452.

\bibitem{yu2007direct}
Z.~Yu, X.~Shao, A direct-forcing fictitious domain method for particulate flows, J. Comput. Phys. 227 (2007) 292--314.

\bibitem{uhlmann2005immersed}
M.~Uhlmann, An immersed boundary method with direct forcing for the simulation of particulate flows, J. Comput. Phys. 209 (2005) 448--476.

\bibitem{tsuji1992lagrangian}
Y.~Tsuji, T.~Tanaka, T.~Ishida, Lagrangian numerical simulation of plug flow of cohesionless particles in a horizontal pipe, Powder Technol. 71 (1992) 239--250.

\bibitem{tsuji1993discrete}
Y.~Tsuji, T.~Kawaguchi, T.~Tanaka, Discrete particle simulation of two-dimensional fluidized bed, Powder Technol. 77 (1993) 79--87.

\bibitem{crowe1977particle}
C.~T. Crowe, M.~P. Sharma, D.~E. Stock, The particle-source-in cell {(PSI-CELL)} model for gas-droplet flows, J. Fluid Eng. 99 (1977) 325--332.

\bibitem{jing2016extended}
L.~Jing, C.~Kwok, Y.~F. Leung, Y.~Sobral, Extended {CFD--DEM} for free-surface flow with multi-size granules, Int. J. Numer. Anal. Methods Geomech. 40 (2016) 62--79.

\bibitem{wu2009three}
C.~Wu, A.~Berrouk, K.~Nandakumar, Three-dimensional discrete particle model for gas--solid fluidized beds on unstructured mesh, Chem. Eng. J. 152 (2009) 514--529.

\bibitem{peng2014influence}
Z.~Peng, E.~Doroodchi, C.~Luo, B.~Moghtaderi, Influence of void fraction calculation on fidelity of {CFD-DEM} simulation of gas-solid bubbling fluidized beds, AIChE J. 60 (2014) 2000--2018.

\bibitem{latzel2000macroscopic}
M.~L{\"a}tzel, S.~Luding, H.~J. Herrmann, Macroscopic material properties from quasi-static, microscopic simulations of a two-dimensional shear-cell, Granul. Matter 2 (2000) 123--135.

\bibitem{gao2021development}
X.~Gao, J.~Yu, L.~Lu, C.~Li, W.~A. Rogers, Development and validation of {SuperDEM-CFD} coupled model for simulating non-spherical particles hydrodynamics in fluidized beds, Chem. Eng. J. 420 (2021) 127654.

\bibitem{wang2022super}
S.~Wang, Y.~Shen, Super-quadric {CFD-DEM} simulation of chip-like particles flow in a fluidized bed, Chem. Eng. Sci. 251 (2022) 117431.

\bibitem{wang2019semi}
Z.~Wang, Y.~Teng, M.~Liu, A semi-resolved {CFD--DEM} approach for particulate flows with kernel based approximation and {Hilbert} curve based searching strategy, J. Comput. Phys. 384 (2019) 151--169.

\bibitem{zhang2023calculation}
Y.~Zhang, W.-L. Ren, P.~Li, X.-H. Zhang, X.-B. Lu, Calculation of particle volume fraction in computational fluid dynamics-discrete element method simulation of particulate flows with coarse particles, Phys. Fluids 35 (2023).

\bibitem{su2025novel}
Z.~Su, C.~Xu, K.~Jia, C.~Cui, X.~Du, A novel semi-resolved {CFD-DEM} coupling method based on point cloud algorithm for complex fluid-particle systems, Comput. Methods Appl. Mech. Eng. 434 (2025) 117561.

\bibitem{deb2013novel}
S.~Deb, D.~K. Tafti, A novel two-grid formulation for fluid--particle systems using the discrete element method, Powder Technol. 246 (2013) 601--616.

\bibitem{che2021novel}
H.~Che, C.~O'Sullivan, A.~Sufian, E.~R. Smith, A novel {CFD-DEM} coarse-graining method based on the {Voronoi tessellation}, Powder Technol. 384 (2021) 479--493.

\bibitem{link2005flow}
J.~Link, L.~Cuypers, N.~Deen, J.~Kuipers, Flow regimes in a spout--fluid bed: A combined experimental and simulation study, Chem. Eng. Sci. 60 (2005) 3425--3442.

\bibitem{kitagawa2001two}
A.~Kitagawa, Y.~Murai, F.~Yamamoto, Two-way coupling of {Eulerian--Lagrangian} model for dispersed multiphase flows using filtering functions, Int. J. Multiph. Flow 27 (2001) 2129--2153.

\bibitem{wang2020semi}
Z.~Wang, M.~Liu, Semi-resolved {CFD--DEM} for thermal particulate flows with applications to fluidized beds, Int. J. Heat Mass Transf. 159 (2020) 120150.

\bibitem{wang2021determination}
Z.~Wang, M.~Liu, On the determination of grid size/smoothing distance in un-/semi-resolved {CFD-DEM} simulation of particulate flows, Powder Technol. 394 (2021) 73--82.

\bibitem{zhu2022semi}
G.~Zhu, Y.~Zhao, Z.~Wang, M.~Liu, et~al., Semi-resolved {CFD-DEM} simulation of fine particle migration with heat transfer in heterogeneous porous media, Int. J. Heat Mass Transf. 197 (2022) 123349.

\bibitem{chen2022semi}
J.~Chen, J.~Zhang, A semi-resolved {CFD-DEM} coupling model using a two-way domain expansion method, J. Comput. Phys. 469 (2022) 111532.

\bibitem{zhang2021optimized}
Y.~Zhang, X.-B. Lu, X.-H. Zhang, An optimized {Eulerian--Lagrangian} method for two-phase flow with coarse particles: {Implementation in} open-source field operation and manipulation, verification, and validation, Phys. Fluids 33 (2021).

\bibitem{sun2015diffusion}
R.~Sun, H.~Xiao, Diffusion-based coarse graining in hybrid continuum--discrete solvers: {Theoretical} formulation and a priori tests, Int. J. Multiph. Flow 77 (2015) 142--157.

\bibitem{capecelatro2013euler}
J.~Capecelatro, O.~Desjardins, An {Euler}--{Lagrange} strategy for simulating particle-laden flows, J. Comput. Phys. 238 (2013) 1--31.

\bibitem{sun2015diffusionapp}
R.~Sun, H.~Xiao, Diffusion-based coarse graining in hybrid continuum--discrete solvers: {Applications} in {CFD--DEM}, Int. J. Multiph. Flow 72 (2015) 233--247.

\bibitem{eshraghi2023coarse}
H.~Eshraghi, E.~Amani, M.~Saffar-Avval, Coarse-graining algorithms for the {Eulerian}-{Lagrangian} simulation of particle-laden flows, J. Comput. Phys. 493 (2023) 112461.

\bibitem{pailha2009two}
M.~Pailha, O.~Pouliquen, A two-phase flow description of the initiation of underwater granular avalanches, J. Fluid Mech. 633 (2009) 115--135.

\bibitem{cheng2021resolved}
K.~Cheng, C.~Zhang, K.~Peng, H.~Liu, M.~Ahmad, Un-resolved {CFD-DEM} method: {An} insight into its limitations in the modelling of suffusion in gap-graded soils, Powder Technol. 381 (2021) 520--538.

\bibitem{zhou2010discrete}
Z.~Zhou, S.~Kuang, K.~Chu, A.~Yu, Discrete particle simulation of particle--fluid flow: model formulations and their applicability, J. Fluid Mech. 661 (2010) 482--510.

\bibitem{anderson1967fluid}
T.~B. Anderson, R.~Jackson, Fluid mechanical description of fluidized beds. {Equations} of motion, Ind. Eng. Chem. Fundam. 6 (1967) 527--539.

\bibitem{cundall1979discrete}
P.~A. Cundall, O.~D. Strack, A discrete numerical model for granular assemblies, Géotechnique 29 (1979) 47--65.

\bibitem{gidaspow1994multiphase}
D.~Gidaspow, Multiphase flow and fluidization: continuum and kinetic theory descriptions, Academic press, 1994.

\bibitem{ergun1952fluid}
S.~Ergun, Fluid flow through packed columns, Chem. Eng. Prog. 48 (1952) 89.

\bibitem{wen1966mechanics}
C.~Y. Wen, Mechanics of fluidization, in: Fluid Particle Technology, Chem. Eng. Progress. Symposium Series, Vol.~62, 1966, pp. 100--111.

\bibitem{liu2022general}
Y.~Liu, X.~Yu, General formulation of drag force on assemblage of spherical particles in fluids: {A} critical review and a new empirical formula, Phys. Fluids 34 (2022).

\bibitem{goldhirsch2010stress}
I.~Goldhirsch, Stress, stress asymmetry and couple stress: from discrete particles to continuous fields, Granul. Matter 12 (2010) 239--252.

\bibitem{weinhart2012discrete}
T.~Weinhart, A.~R. Thornton, S.~Luding, O.~Bokhove, From discrete particles to continuum fields near a boundary, Granul. Matter 14 (2012) 289--294.

\bibitem{cheng2023concurrent}
H.~Cheng, A.~R. Thornton, S.~Luding, A.~L. Hazel, T.~Weinhart, Concurrent multi-scale modeling of granular materials: {Role} of coarse-graining in {FEM-DEM} coupling, Comput. Methods Appl. Mech. Eng. 403 (2023) 115651.

\bibitem{weinhart2016influence}
T.~Weinhart, C.~Labra, S.~Luding, J.~Y. Ooi, Influence of coarse-graining parameters on the analysis of {DEM} simulations of silo flow, Powder Technol. 293 (2016) 138--148.

\bibitem{wang2024computational}
S.~Wang, S.~Ji, Computational mechanics of arbitrarily shaped granular materials, Springer, 2024.

\bibitem{chukkapalli1999scheme}
G.~Chukkapalli, S.~R. Karpik, C.~R. Ethier, A scheme for generating unstructured grids on spheres with application to parallel computation, J. Comput. Phys. 149 (1999) 114--127.

\bibitem{baselga2018fibonacci}
S.~Baselga, Fibonacci lattices for the evaluation and optimization of map projections, Comput. Geosci. 117 (2018) 1--8.

\bibitem{swinbank2006fibonacci}
R.~Swinbank, R.~James~Purser, Fibonacci grids: {A} novel approach to global modelling, Q. J. R. Meteorol. Soc. 132 (2006) 1769--1793.

\bibitem{saff1997distributing}
E.~B. Saff, A.~B. Kuijlaars, Distributing many points on a sphere, Math. Intelligencer 19 (1997) 5--11.

\bibitem{gonzalez2010measurement}
{\'A}.~Gonz{\'a}lez, Measurement of areas on a sphere using {Fibonacci} and latitude--longitude lattices, Math. Geosci. 42 (2010) 49--64.

\bibitem{ten2002particle}
A.~Ten~Cate, C.~Nieuwstad, J.~J. Derksen, H.~Van~den Akker, Particle imaging velocimetry experiments and lattice-boltzmann simulations on a single sphere settling under gravity, Phys. Fluids 14 (2002) 4012--4025.

\bibitem{xie2021cfd}
Z.~Xie, S.~Wang, Y.~Shen, {CFD-DEM} modelling of the migration of fines in suspension flow through a solid packed bed, Chem. Eng. Sci. 231 (2021) 116261.

\bibitem{guo2011motion}
J.~Guo, Motion of spheres falling through fluids, J. Hydraul. Res. 49 (2011) 32--41.

\bibitem{khan2016pressure}
M.~S. Khan, S.~Mitra, S.~Ghatage, Z.~Peng, E.~Doroodchi, B.~Moghtaderi, J.~B. Joshi, G.~M. Evans, Pressure drop and voidage measurement in solid-liquid fluidized bed: {Experimental}, mathematical and computational study, Chemeca 2016: Chemical Engineering-Regeneration, Recovery and Reinvention (2016) 1019--1030.

\bibitem{iverson2000acute}
R.~M. Iverson, M.~Reid, N.~R. Iverson, R.~LaHusen, M.~Logan, J.~Mann, D.~Brien, Acute sensitivity of landslide rates to initial soil porosity, Science 290 (2000) 513--516.

\bibitem{yang2020pore}
G.~Yang, L.~Jing, C.~Kwok, Y.~D. Sobral, Pore-scale simulation of immersed granular collapse: implications to submarine landslides, J. Geophys. Res. Earth Surf. 125 (2020) e2019JF005044.

\bibitem{xiao2011algorithms}
H.~Xiao, J.~Sun, Algorithms in a robust hybrid {CFD-DEM} solver for particle-laden flows, Commun. Comput. Phys. 9 (2011) 297--323.

\bibitem{ries2014coarse}
A.~Ries, L.~Brendel, D.~E. Wolf, Coarse graining strategies at walls, Comput. Part. Mech. 1 (2014) 177--190.

\bibitem{zhu2002averaging}
H.~Zhu, A.~Yu, Averaging method of granular materials, Phys. Rev. E 66 (2002) 021302.

\bibitem{kloss2012models}
C.~Kloss, C.~Goniva, A.~Hager, S.~Amberger, S.~Pirker, Models, algorithms and validation for opensource {DEM} and {CFD--DEM}, Prog. Comput. Fluid Dyn. 12 (2012) 140--152.

\bibitem{radl2015state}
S.~Radl, B.~C. Gonzales, C.~Goniva, S.~Pirker, State of the art in mapping schemes for dilute and dense euler-lagrange simulations, in: Selected papers from 10th International Conference on Computational Fluid Dynamics in the Oil \& Gas, Metallurgical and Process Industries, SINTEF Academic Press, 2015.

\bibitem{goniva2015open}
C.~Goniva, B.~Blais, S.~Radl, C.~Kloss, Open source {CFD-DEM} modelling for particle-based processes, in: 11th International Conference on Computational Fluid Dynamics in the Minerals and Process Industries, 2015.

\end{thebibliography}

\end{document}